\documentclass[sigconf]{acmart}

\usepackage{subcaption}
\usepackage{enumitem} 
\usepackage{colortbl}
\usepackage{booktabs} 
\usepackage{pifont} 
\usepackage{appendix}

\AtBeginDocument{%
  }

\copyrightyear{2025}
\acmYear{2025} 
\setcopyright{cc}
\setcctype{by}
\acmDOI{10.1145/3706598.3713716}

\acmConference[CHI '25]{CHI Conference on Human Factors in Computing Systems}{April 26-May 1, 2025}{Yokohama, Japan}

\acmBooktitle{CHI Conference on Human Factors in Computing Systems (CHI'25), April 26-May 1, 2025, Yokohama, Japan}
\acmISBN{979-8-4007-1394-1/25/04}

\newcommand{\mbd}{\textsc{Media Bias Detector}}

\begin{document}

\title[\mbd]{\mbd: Designing and Implementing a Tool for Real-Time Selection and Framing Bias Analysis in News Coverage}

\author{Jenny S Wang}
\authornote{Both authors contributed equally to this research.}
\orcid{0009-0004-9964-8954}
\affiliation{%
  \institution{Harvard Business School}
  \city{Boston}
  \state{Massachusetts}
  \country{USA}
}
\email{jewang@hbs.edu}

\author{Samar Haider}
\authornotemark[1]
\orcid{0009-0000-5725-8887}
\affiliation{%
  \institution{University of Pennsylvania}
  \city{Philadelphia}
  \state{Pennsylvania}
  \country{USA}
}
\email{samarh@seas.upenn.edu}

\author{Amir Tohidi}
\orcid{0000-0003-4572-9834}
\affiliation{%
  \institution{University of Pennsylvania}
  \city{Philadelphia}
  \state{Pennsylvania}
  \country{USA}
}
\email{atohidi@seas.upenn.edu}

\author{Anushkaa Gupta}
\orcid{0009-0002-0743-5712}
\affiliation{%
  \institution{University of Pennsylvania}
  \city{Philadelphia}
  \state{Pennsylvania}
  \country{USA}
}
\email{agupta08@sas.upenn.edu}

\author{Yuxuan Zhang}
\affiliation{%
  \institution{University of Pennsylvania}
  \city{Philadelphia}
  \state{Pennsylvania}
  \country{USA}
}
\orcid{0009-0002-0743-5712}
\email{yuxuanzh@seas.upenn.edu}

\author{Chris Callison-Burch}
\orcid{0000-0001-8196-1943}
\affiliation{
  \institution{University of Pennsylvania}
  \city{Philadelphia}
  \state{Pennsylvania}
  \country{USA}
}
\email{ccb@upenn.edu}

\author{David Rothschild}
\orcid{0000-0002-7792-1989}
\affiliation{
  \institution{Microsoft Research}
  \city{New York}
  \state{New York}
  \country{USA}
}
\email{david@researchdmr.com}

\author{Duncan J Watts}
\orcid{0000-0001-5005-4961}
\affiliation{
  \institution{University of Pennsylvania}
  \city{Philadelphia}
  \state{Pennsylvania}
  \country{USA}
}
\email{djwatts@seas.upenn.edu}

\renewcommand{\shortauthors}{Wang et al.}

\begin{abstract}
Mainstream media, through their decisions on what to cover and how to frame the stories they cover, can mislead readers without using outright falsehoods. Therefore, it is crucial to have tools that expose these editorial choices underlying media bias. In this paper, we introduce the \mbd, a tool for researchers, journalists, and news consumers. By integrating large language models, we provide near real-time granular insights into the topics, tone, political lean, and facts of news articles aggregated to the publisher level. We assessed the tool’s impact by interviewing 13 experts from journalism, communications, and political science, revealing key insights into usability and functionality, practical applications, and AI's role in powering media bias tools. We explored this in more depth with a follow-up survey of 150 news consumers. This work highlights opportunities for AI-driven tools that empower users to critically engage with media content, particularly in politically charged environments. 

\end{abstract}

\ccsdesc[500]{Human-centered computing~User studies}
\ccsdesc[500]{Human-centered computing~Empirical studies in HCI}
\ccsdesc[300]{Information systems~Web applications}

\keywords{media bias, news analysis, large language models (LLMs), LLM-driven tools}


\maketitle

\section{Introduction}
\label{sec:intro}

Since the 2016 election, there has been growing concern about the pervasive impact of fake news \cite{lazar_2018, soroush_2018, pennycook2021psychology, allcott2017social}. However, in practice, most people consume news from mainstream sources, where stories, though factually accurate, can still be \emph{biased} \cite{allen2020evaluating}. This bias arises from the considerable control that journalists and editors exert over \emph{selection} (choosing to emphasize some issues, events, or people over others) as well as their \emph{framing} (choosing the tone, perspective, or facts of a story to present an issue in a particular way) \cite{entman1993framing,  chong2007framing, Rothschild2023}. Unlike outright falsehoods, selection and framing can subtly mislead readers, making these forms of bias difficult for traditional fact-checking tools to detect \cite{park2021presence}. Experimental evidence has shown that it is remarkably easy to mislead a reader without making any explicitly false statements \cite{alexander2022media, lifchits2021success}.  Moreover, because factually accurate but biased news coverage tends to be consumed by much larger populations than categorically fake news, its negative impact can be considerably larger in aggregate \cite{allen2024quantifying}.

Measuring and exposing media bias, which we define as the preferential selection of some stories, facts, people, events, or perspectives over others, is critical to challenging the illusion of objectivity promoted by many news organizations. By revealing how editorial choices (decisions about which facts, voices, and perspectives to emphasize or downplay) shape coverage, readers can better understand both the context of the information they consume and the perspectives they are missing. A lack of diverse perspectives can contribute to echo chambers, where individuals relying on a single source or a limited set of ideologically similar sources are more likely to encounter “separate realities,” fostering disagreement on basic facts or priorities \cite{muise_etal, hosseinmardi2023diminishing, bram2024beyond}.

Communications researchers have long emphasized the importance of  media bias \cite{stroud2008media, baum2008new,chong2007framing, xiang2007news, chiang2011media}; however, methodological limitations have until recently prevented its identification at scale. Traditional approaches, like manually quantifying topics, viewpoints, or facts across publishers, are costly and time-consuming \cite{budak2016fair,lim2020annotating, mitchell2017covering}, while natural language processing (NLP) methods such as n-gram counting typically fail to capture the full context of an article \cite{gentzkow2010drives, d2000media}. Now, with advancements in large language models (LLMs), it has become possible to annotate large documents faster and more efficiently, while maintaining high accuracy \cite{goel2023llms}. Specifically, LLMs can efficiently ingest and parse tens of thousands of articles, extracting key features such as topics, subtopics, and facts as well as labeling them according to subjective quantities such as tone (positive vs. negative) and partisanship (Democrat vs. Republican) \cite{pena2023leveraging, xu2023large}. This ability to extract nuanced information from large datasets makes LLMs a valuable tool for identifying and analyzing media bias. 

In this paper, we present the \mbd, an LLM-driven tool designed to dynamically track and analyze major news coverage from a diverse collection of prominent publishers across the political spectrum.\footnote{We limit our initial investigation to ten mainstream news publishers to maintain data quality and keep labeling costs manageable. But, the tool is easily adaptable to additional publishers.} We operationalize media bias in two distinct ways, corresponding to the aforementioned distinction between selection and framing bias. First, we quantify selection bias by measuring the differential attention paid to different news categories (e.g. politics vs. business) as well as topics (e.g. the 2024 election within politics) and subtopics (e.g. the presidential horse race within the 2024 election). Second, we quantify framing bias along two key dimensions: a) political lean, which we define as the extent to which an article aligns with the viewpoints, policies, or concerns of Republicans versus Democrats, either explicitly or implicitly; and b) tone, defined as the emotional slant of coverage (i.e. positive, neutral, negative). Unlike existing tools, which typically consider subsets of news stories and label publishers broadly as left, right, or center, the \mbd\ annotates individual articles produced every day. These annotations are then aggregated to provide a dynamic, data-driven view of each publisher, reflecting the diversity of their coverage rather than assigning static ideological labels. Our inclusion of tone also addresses a gap in media bias detection, which often focuses on political lean but has paid less attention to sentiment as a source of bias \cite{robertson2023negativity}. This added granularity allows readers to explore the distribution of lean and tone across articles published by a news outlet. For instance, a user might find that while the Wall Street Journal appears generally neutral overall, its coverage leans more Republican on immigration and slightly more Democratic on the environment and reproductive rights.

The \mbd\ allows users---whether researchers, journalists, or everyday news consumers---to interactively explore and quantify media bias by examining the coverage focus, article type (i.e., report, analysis, or opinion), tone, and political lean of the top news stories. With this tool, users can examine how different publishers emphasize or ignore different topics, such as whether they focus more on Joe Biden's or Donald Trump's age, or whether they frame the election as an entertaining horse race versus a discussion of key policy differences. It also allows users to explore topic-specific questions, such as quantifying how much inflation is discussed compared to wages, or how crime is framed: does it focus on the perspective of law enforcement, economic impact, or the effects of incarceration? Furthermore, the \mbd\ helps users keep up with the fast pace of the news cycle by summarizing the top events of the day and highlighting key snippets of information from these stories. To evaluate the effectiveness of the \mbd\ and inform future improvements, we conducted semi-structured interviews with 13 experts in media, communications, and politics, as well as a survey of 150 everyday news consumers. The goal of collecting feedback from these user groups was to address the following research questions:

\begin{enumerate}[label=\textbf{RQ\arabic*.}, leftmargin=*, labelindent=0pt]
    \item Can we develop a tool that effectively conveys media bias to users while remaining easy to use? 
    \item How does the use of LLMs in the \mbd\ influence users' trust in the tool compared to human-generated ratings?
    \item Which audiences are most likely to benefit from the \mbd, and how can the tool be optimized to better serve these groups?
\end{enumerate}

In our evaluation of the \mbd, several key themes emerged. Experts appreciated the \mbd's ability to consolidate complex information and provide multiple forms of evaluation, emphasizing its value for media literacy education and qualitative research. They also stressed the importance of transparency when using LLMs to classify information related to complex topics like media bias. Everyday users found the tool useful and accessible, often discovering new insights they had not known before, and generally placed more trust in AI tools than experts. Both groups, however, valued the reassurance of human oversight. Overall, the \mbd\ proved to be a versatile tool that appealed to a wide range of audiences for education, research, and daily bias checks.

To summarize, the main contributions of our work are as follows: 

\begin{enumerate} 
    \item The implementation of the \mbd, an interactive online tool that aggregates key information from news articles and guides users to explore different facets of media bias. 
    \item The results of a user study with 13 media experts and a survey of 150 news consumers, which demonstrate the \mbd's broad applicability and value in helping users understand and discover media bias in online news. \end{enumerate}

\section{Background and Related Work}
\label{sec:related_work}

Bias, in statistical theory, implies the presence of a "true" value from which the biased estimate systematically differs. However, in practice, we generally lack a ground truth, making it infeasible to directly assess bias for individual articles or even entire publishers. This assertion might seem counterintuitive, as we often perceive articles as biased when we read them. This perception arises because we implicitly treat our subjective opinions as the "truth," even though they are likely biased as well \cite{pronin2004objectivity}. In the absence of a universally accepted ground truth, we instead need to identify bias by observing patterns and differences in how news is reported between publishers, or within publishers over time. In this section, we anchor our discussion of media bias in the concepts of selection and framing biases and consider their relevance to designing media bias tools grounded in HCI principles. We also situate the \mbd\ among existing media bias detection tools and discuss their limitations. Finally, we review current literature on LLMs for annotation tasks and explain why these models are well-suited for analyzing media bias. 

\subsection{Agenda-Setting, Framing, and the Role of HCI in Designing Tools for Media Bias}
\label{sec:related_work_media_bias}

Media bias presents major implications for the HCI community as online platforms increasingly become the primary medium for news consumption. Within the communications field, media bias is often analyzed through two key theories: agenda-setting and framing. Agenda-setting theory argues that the media doesn't tell us what to think but rather what to think about \cite{agenda_setting_1972, mccombs2005agenda}. The process of \emph{selection bias}---choosing certain stories or facts to report while omitting others---directly contributes to the media's agenda-setting power, shaping public discourse by directing attention to specific topics over others \cite{dearing1996agenda}. For example, a news outlet might publish many stories on a frontrunner's campaign rallies while giving limited coverage to other candidates. Agenda-setting theory is highly relevant to HCI because digital platforms increasingly mediate what content users see \cite{Naser2020RelevanceAC}. Algorithms, interface design, and personalized recommendation systems often act as agenda-setters, determining what content is presented or excluded from users' screens \cite{maudet,Eslami2016FirstI}. This phenomenon raises critical questions in HCI about how systems shape user awareness and understanding, as well as how tools can be designed to challenge these embedded biases.

Beyond selecting which issues to emphasize, the media also influences public perception via the manner in which information is presented---a process known as framing \cite{weaver2007thoughts}. \emph{Framing bias} builds on selection bias by not just determining what is covered, but how it is covered. This involves the inclusion or omission of specific details and perspectives, the tone or language used, the context provided, or the omission of background information, all of which can significantly alter how an issue is perceived by the audience \cite{entman1993framing,gentzkow2006media, chong2007framing}. For instance, framing a protest as a "riot" in one article versus a "peaceful demonstration" in another can significantly influence how readers perceive the event \cite{brown_protest_2021, susanszky2022media}. Similarly, framing immigration as an "opportunity for economic growth" versus an "immigration crisis" can sway public perception on the issue \cite{Igartua2009ModeratingEO}. Framing effects relate to HCI through a shared focus on how the presentation and emphasis of information influences its importance or impact \cite{hartmann_framing}. Media bias tools informed by HCI principles can play a critical role in revealing framing biases \cite{dingler_workshop}. For example, tools that allow users to analyze data from multiple perspectives \cite{ainsworth2008educational}, explore specific points in more detail \cite{springer2019progressive}, or compare diverse sources \cite{bhuiyan2023newscomp,braaten2011measuring} can foster a deeper understanding of how the same event is portrayed in different ways.

Although the concepts of selection and framing bias are well-known in media studies and communications \cite{o2009news, d2000media, matthes2009s}, their impact has become even more pronounced in today's digital age \cite{harcup2017news, bourgeois2018selection}. The rapid dissemination and broad reach of online news allow information consumers to quickly access content that reinforces their pre-existing opinions. Research indicates that online news consumption exhibits a polarized pattern, with users spending significantly more time on news sources that align with their political leanings compared to those that do not \cite{garimella2021political}. Tools that make media bias visible---especially in terms of selection and framing biases---can help users recognize these patterns and engage more critically with their news consumption \cite{markus}. By applying HCI principles to these tools, we can design interfaces that promote media literacy and encourage balanced engagement with digital news platforms.

\subsection{Examining Current Approaches and Limitations in Assessing Media Bias}
\label{sec:related_work_existing_tools}

Scholars have proposed numerous taxonomies to understand media bias, yet no universally accepted set of media bias metrics or standard measurements exist \cite{puglisi2015empirical, gentzkow2015media, morstatter_2018, hamborg2019automated, huang2024uncovering}. Current methods often reduce news publishers to a single metric by assigning a political lean rating from "Left" to "Right" \cite{allsides,mediabiasfactcheck,groundnews}. These tools simplify bias labels to make it easier for users to digest, such as using 2-dimensional axes with political lean and reliability \cite{adfontesmedia} or a 1-dimensional categorization of publishers across the political spectrum \cite{allsides}. Although useful, these static classifications fail to capture within-publisher differences in bias over time, across topics, or even across individual articles. 

Evaluating bias at the article level is less common and more challenging. Earlier tools evaluating article level bias relied on basic NLP techniques, such as keyword frequency analysis, to cluster similar stories in the media \cite{park2009newscube}. However, these methods are limited in their ability to account for context and the evolving nature of news stories. Recent advancements in LLMs offer improved contextual understanding capabilities which can more accurately help identify media bias in news coverage. Even though other proprietary tools that incorporate AI features have emerged \cite{biasly,allsides}, these platforms generally do not update to reflect media bias in recent news coverage. Furthermore, the lack of transparency around their AI systems makes it unclear which models they use and how they arrive at their conclusions.

The \mbd\ addresses these limitations by moving beyond oversimplified metrics and publisher level analyses. Rather than simplifying media bias to one or two dimensions, the \mbd\ captures framing and selection biases through a multidimensional analysis of publisher coverage over time. This approach allows users to compare tone, political lean, and content focus between publishers, and across topics, resulting in a more comprehensive and dynamic approach to media bias. The \mbd\ is intentionally designed to provide near real-time updates, allowing users to explore how media coverage---and the biases within it---shift dynamically as new information emerges. Furthermore, we make our underlying methodology, including our model usage, prompt phrasing, and human in the loop verification framework, readily available to all users of our tool (see Appendix \ref{sec:website_methodology}).

\subsection{Incorporating Large Language Models (LLMs) for Media Bias Detection}
\label{sec:relate_work_llm_bias}

LLMs, such as GPT-4, utilize deep contextual embeddings to capture subtle semantic relationships and nuances within text \cite{brown2020language,achiam2023gpt, radford2019language}. Studies have shown that LLMs perform well in generative tasks such as summarization \cite{liu-etal-2024-learning, ravaut-etal-2024-context, liu2023learning}, as well as discriminative tasks like sentiment analysis and topic classification \cite{chang2024survey, pena2023leveraging, zhang2023sentiment}. More recent studies have shown that state-of-the-art LLMs exhibit increasingly complex reasoning and problem-solving capabilities \cite{bubeck2023sparks}. These findings suggest that LLMs can be especially useful for the complex task of media bias detection, where understanding context, tone, and subtle language is crucial.

While traditional machine learning models used in annotation tasks require task-specific training, instruction-tuned pre-trained LLMs can generalize to different tasks in a zero-shot setting \cite{wang_humanllm_2024}. Prior work has shown that using LLMs to automate the initial stages of data annotation enables researchers to quickly process large volumes of content that would be unmanageable with manual annotation alone \cite{ziems2024can}. This scalability is critical given the speed with which major news outlets publish new articles, which must be analyzed in near real-time to capture evolving narratives and help readers interpret the news. By incorporating LLMs into the \mbd's news analysis pipeline, we can efficiently extract detailed information like sentence-level codings and classifications by topic, subtopic, article type, tone, and political lean, while significantly increasing the volume of articles we can annotate. However, it is important to acknowledge potential limitations of LLMs in zero-shot settings, such as political biases in their outputs and the influence of this bias on the tool's results. These challenges are explored in greater depth in Section \ref{sec:limitations}.

The integration of LLMs into traditionally human processes is informed by a growing body of research that shows that human-AI collaboration enhances the accuracy and reliability of automated systems \cite{uchendu2023does, goel2023llms}. Although LLMs can quickly process large datasets and perform initial classifications, human oversight is crucial for providing context that LLMs do not have for current events and verifying the model's outputs for consistent standards of accuracy \cite{wang_humanllm_2024, amirizaniani2024developing}. In the \mbd, humans play an important role at every step, from generating a targeted list of relevant news topics for LLMs to classify, to continuously monitoring the model's classifications over time.

We intentionally chose to evaluate the responses generated by LLMs through a process of human annotation that emphasizes validation over independent labeling. Rather than having annotators blindly label the data, we engage them in a validation task where they read GPT-generated responses to assess their coherence and soundness. Although having humans independently label content and then comparing it to GPT's output could yield interesting insights, it is important to recognize that human disagreements often occur, even among reasonable individuals. Particularly in complex tasks, such as reading an entire article and labeling its lean and tone, as well as extracting facts, achieving high agreement is challenging \cite{mitchell2018distinguishing}. Our methodology accounts for this inherent subjectivity and aims to ensure that topics presented on the \mbd\ are coherent, reasonable, and free from overt inaccuracies. By involving humans in the validation process, we impose a layer of quality control that acknowledges the absence of a singular ground truth. 

\section{Design Considerations}
\label{sec:design_considerations}

The overarching design goal of our tool is not to determine the media bias of each publisher for the user, but to provide affordances that allow users to break down information and compare mainstream media publications. These affordances enable users to see the aggregated raw data broken down by topic and subtopic, and to compare the overall lean and tone within each topic and subtopic to understand the nature of each publisher's views. This section outlines our key design considerations, summarized in Table \ref{tab:design_considerations}.

\subsection*{D1: Enable \emph{broad} exploration of the same data.}
\label{sec:design_considerations_d1}

Media bias is a multidimensional problem \cite{eberl2017one} and cannot accurately be captured by just one metric. A central goal of our tool is to support \emph{breadth} of exploration, providing users with affordances that enable them to examine the data from multiple perspectives and draw their own conclusions. This design is informed by principles of learning with multiple representations, which enhance cognitive processes in learning by allowing users to approach complex data from different angles \cite{ainsworth2008educational, ainsworth1999functions}. To operationalize these principles, we use faceted categories to structure the interface, enabling users to dynamically toggle between variables and uncover relationships that might otherwise be difficult to identify \cite{hearst_faceted,lee_faceted}. For example, users can switch between metrics such as coverage volume, political lean, and tone across publishers and topics. They can also adjust the date range to compare publishers over any time frame and filter by article type, allowing for a comprehensive comparison of how various content categories reflect media bias over selected time periods. Figures \ref{fig:coverage_bar} and \ref{fig:coverage_grid} showcase how the \mbd\ incorporates faceted categories to help users process information through multiple representations.

\subsection*{D2: Enable \emph{deep} exploration of specific data.}
\label{sec:design_considerations_d2}

In addition to being a multidimensional problem, media bias is also a \textit{multilevel} one. While overall news coverage on a particular category may not differ much between two publishers, how they choose to cover specific topics, subtopics, and events within that category can reveal a hidden bias. A major affordance of our tool is the ability to engage in \textit{deep} exploration of the data after seeing an overview, allowing users to uncover more detailed biases without being overwhelmed at the very first stage. This approach, informed by the principle of \emph{progressive disclosure}, affords a gradual reveal of complexity as users build familiarity with the system \cite{carroll_1984, smith1982star, nakatani1983soft, springer2019progressive}. 
From a media bias standpoint, we apply progressive disclosure by allowing users to begin with broad categories and gradually narrow their focus to more specific areas. As users engage with a particular category, they can drill down into increasingly detailed topics and subtopics, enabling a deeper exploration of media coverage. This also extends beyond topic-level analysis, as our tool affords sentence-level text annotation which helps users examine how different news publishers write about the same events. Users can begin by exploring a news event cluster, dive into the top facts reported about that event, and then compare how the same facts are presented across different publications. Figure \ref{fig:depth} illustrates how a user can progressively refine their analysis at the category, topic, and subtopic levels.

\begin{table*}[t]
\centering
\caption{Design considerations for the \mbd, summarizing the goals, descriptions, and key strategies employed to address media bias.}
\begin{tabular}{p{0.18\linewidth} p{0.26\linewidth} p{0.48\linewidth} }
\hline
\textbf{Design Consideration} & \textbf{Description} & \textbf{Key Features and Strategies} \\
\hline
\textbf{D1}: Broad exploration of the same data & Media bias is a multidimensional problem that requires users to explore data from various perspectives. & - Facilitates multiple representations of data to enhance cognitive understanding. \newline - Implements faceted categories to allow toggling between metrics such as coverage volume, political lean, and tone. \\

\textbf{D2}: Deep exploration of specific data & Media bias is a multilevel problem, requiring detailed examination beyond overviews to reveal nuanced biases. & - Uses progressive disclosure to reduce cognitive load and gradually reveal complexity. \newline - Allows drilling down into detailed topics, subtopics, and events within broader categories. \newline - Provides sentence-level text annotation for event-level comparisons.  \\

\textbf{D3}: Enable easy comparison across publishers and topics & Media bias can be more clearly understood by comparing similarities and differences between publishers and topics. & - Designed to expose key elements of news stories across the political spectrum using side-by-side comparisons. \newline - Integrates breadth (D1) and depth (D2) principles for cohesive comparative analysis. \\
\hline
\end{tabular}
\label{tab:design_considerations}
\end{table*}

\subsection*{D3: Enable easy comparison across publishers and topics.}
\label{sec:design_considerations_d3}

News readers often have limited time and tend to rely on a single source or ideologically aligned sources to learn about current events \cite{iyengar2009red}. The affordance of comparative analysis within our tool allows users to easily identify similarities and differences between multiple publishers in one place, which may help them recognize discrepancies between perceived and actual coverage \cite{bhuiyan2023newscomp, stahl1996happens, braaten2011measuring}. To reduce cognitive load, our tool provides side-by-side comparisons of publishers and topics with the goal of better contextualizing the broader news ecosystem without users needing to switch between sources. 

While D1 emphasizes breadth by encouraging users to explore data across multiple dimensions and D2 enables deep, focused analysis, D3 integrates these principles into a cohesive framework for comparison. Table \ref{tab:design_considerations} provides a summary of these design considerations and key features addressed by each of them. We used these considerations to design and implement the \mbd, an interactive dashboard that exposes and aggregates key elements of top news stories across the political spectrum in close to real time. Throughout Section \ref{sec:design}, we illustrate how the \mbd\ facilitates comparative analysis through its various dashboards and visualization formats, helping users to critically evaluate similarities and differences across publishers and topics.

\section{\mbd: An Interactive Near-Real-Time News Tracking Tool}
\label{sec:design}

This section introduces the design of the \mbd, a tool built to help users navigate the complexities of media bias through three guiding principles outlined in Section \ref{sec:design_considerations}: broad exploration (\textbf{D1}), deep and multilevel exploration (\textbf{D2}), and comparative analysis (\textbf{D3}).\footnote{From this point forward, we refer to these design considerations as D1, D2, and D3 for clarity and consistency.} The dashboard currently focuses on ten prominent online news publishers, selected for their mix of reach and agenda-setting influence: Associated Press News, Breitbart News, CNN, Fox News, The Guardian, The Huffington Post, The New York Times, USA Today, The Wall Street Journal, and The Washington Post. Although this focus ensures a balanced representation of diverse viewpoints, it is important to note the inherent limitation of restricting the news articles to only ten sources, which may exclude other influential or regional publishers. We discuss this limitation further in Section \ref{sec:limitations}.

The user interface (UI) features two primary views: the \emph{Coverage dashboard}, which visualizes the volume, lean, and tone of articles published on different topics; and the \emph{Events dashboard}, which tracks the top events covered by publishers in the preceding three days and highlights their top facts. Events are defined as significant incidents that generate a large number of articles within a short time frame. We detect them algorithmically by clustering similar articles published within a one-day window.\footnote{Initially, a three-day window was used, but this was later adjusted to one day to better capture emerging events in today's fast-paced news environment.} The underlying mechanics of the \mbd\ rely on large-scale automated analysis, using OpenAI's GPT-4o with human-in-the-loop review to classify articles by topic, subtopic, article type, tone, and political lean. To accommodate the scale of the data and the need to efficiently run analyses every day, we use AWS S3 for storage and utilize OpenAI's asynchronous API to process requests in parallel and achieve maximum throughput within existing rate limits. Additional methodological details are provided in Appendix \ref{sec:website_methodology}. Here, we examine how users interact and engage with the design of an LLM-driven media bias tool by focusing on the features and affordances users find most valuable (RQ1), how the awareness of LLMs being used affects their trust in the tool (RQ2), and which user groups benefit most from using the tool's more sophisticated set of features (RQ3). 

\subsection{Coverage Dashboard}
\label{sec:design_coverage}

\begin{figure*}
    \centering
    \includegraphics[width=0.95\textwidth]{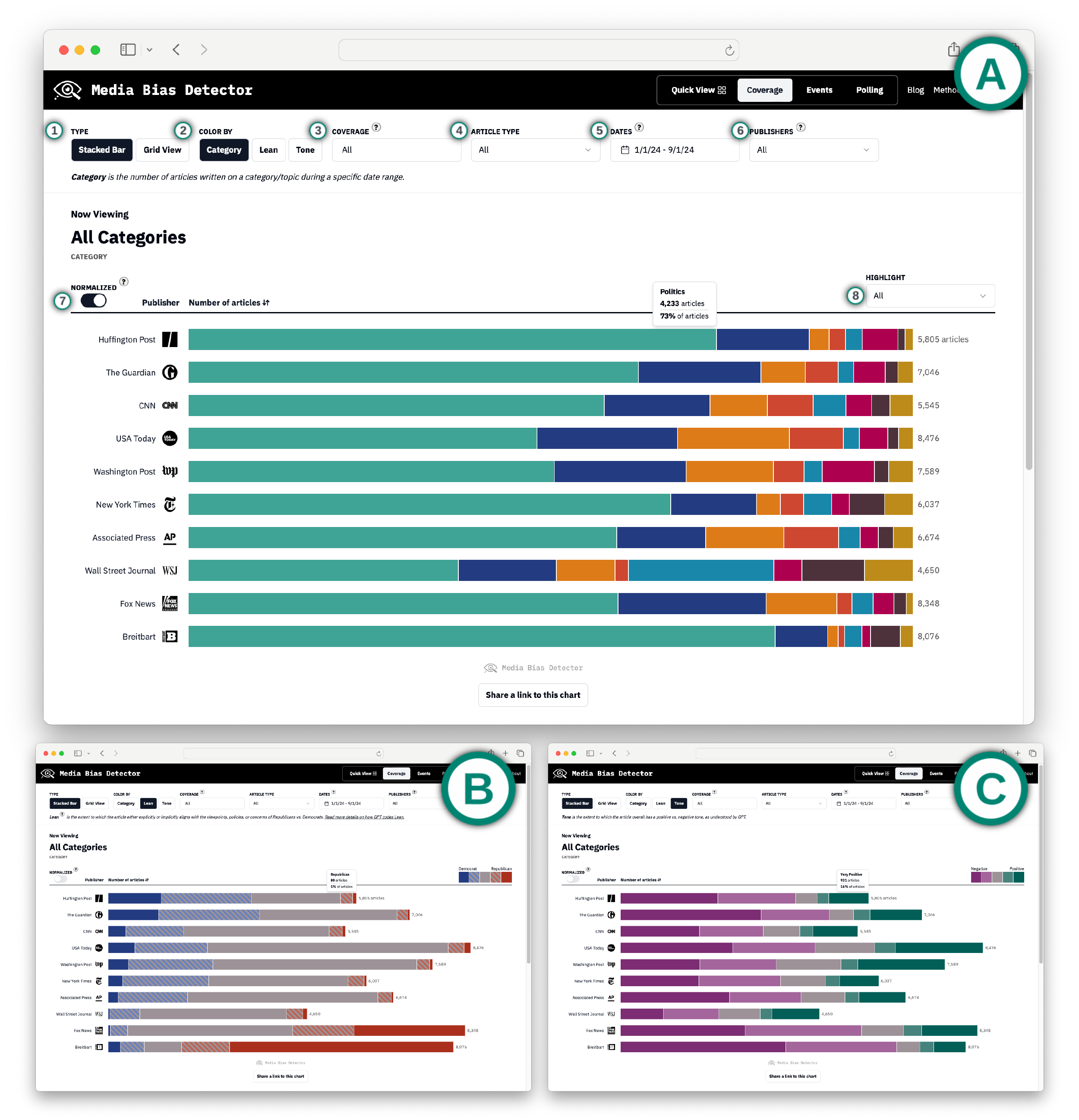}
    \caption{The default view of the Coverage dashboard which allows \textit{broad} exploration of the date \textbf{(D1)}. Every user lands at a screen showing the category-wise coverage of the news publishers (A). Each publisher is represented with a stacked bar where a segment represents the number of articles published by them in a given category, allowing them to directly compare the proportion of attention they give to those topics \textbf{(D3)}. Users are provided with a variety of controls to adjust the chart type and color, and to filter on publishers, article type, and date range \textbf{(D1)}. Users can also toggle normalization off to allow direct comparisons of absolute numbers instead of the proportion of coverage. Coloring by Lean (B) shows the distribution of articles that are more aligned with a given political viewpoint, and coloring by Tone (C) shows the variation in sentiment across the same articles. Hovering on a segment displays a tooltip explaining what it represents and presenting the count and proportion of articles that fall within it.}
    \label{fig:coverage_bar}
\end{figure*}

\begin{figure*}
    \centering
    \includegraphics[width=\textwidth]{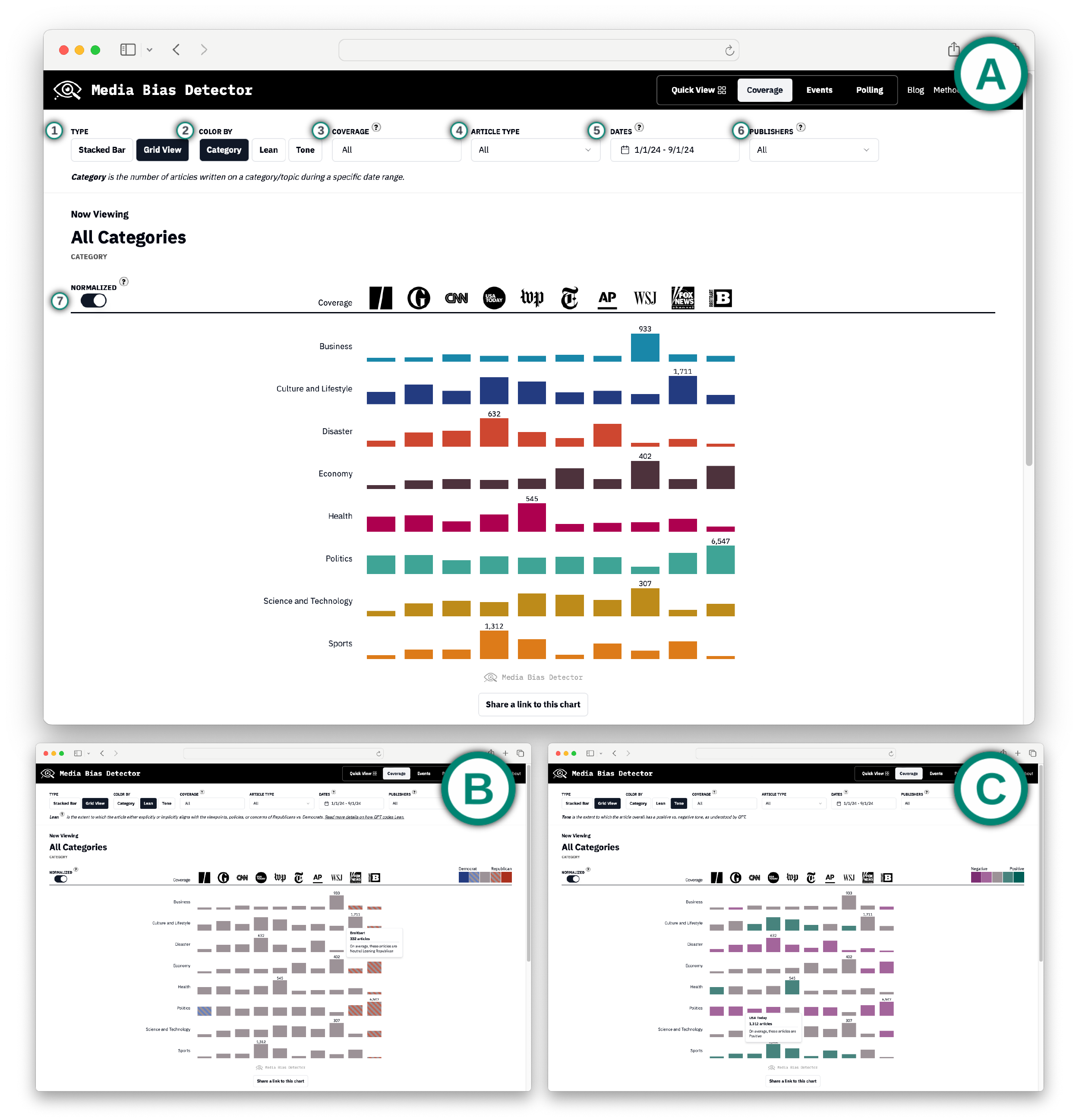}
    \caption{The grid view of the Coverage dashboard presents an alternative visualization to the stacked bar in Figure \ref{fig:coverage_bar} by giving each bar segment its own cell in a grid. This allows similarly \textit{broad} exploration (\textbf{D1}) by enabling more direct comparisons between different categories and publishers (\textbf{D3}) without the need to hover or click on a news category to highlight it. Each cell shows how many articles were published on a given topic by a particular publisher, with the highest-publishing publisher highlighted by its article count. The news category color map for this view matches that of the stacked bar chart. Similar to that view, the bars can be colored by Lean (B) or Tone (C), where the cell color represents the average political lean or tone of the articles in that category. Hovering on a cell displays a tooltip explaining what it represents and presenting the count of articles that fall within it.}
    \label{fig:coverage_grid}
\end{figure*}

\begin{figure*}
    \centering
    \includegraphics[width=0.8\textwidth]{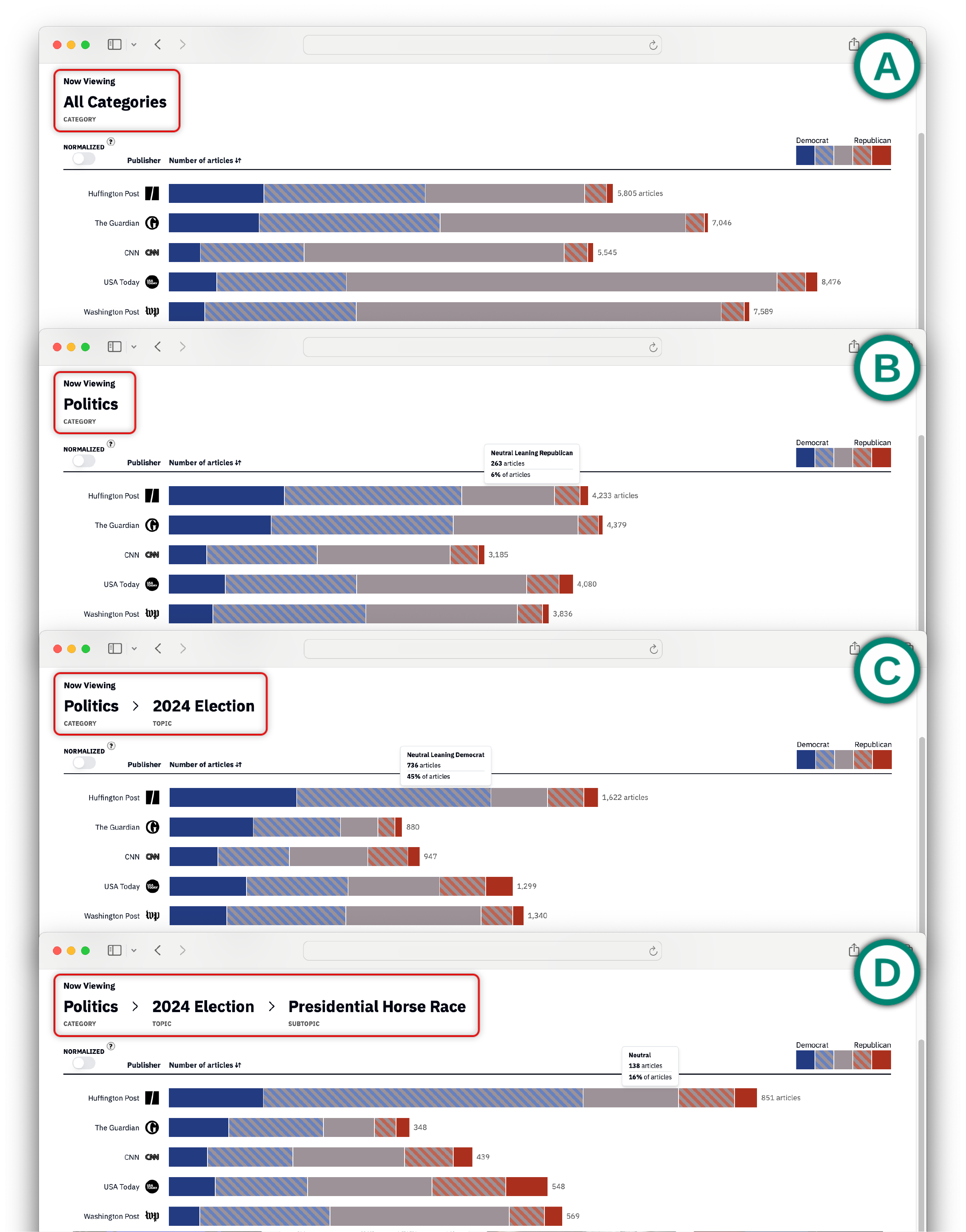}
    \caption{To enable \textit{deep} exploration of specific data (\textbf{D2}), we allow users to click through a hierarchy of news topics and subtopics to zoom into news of interest to them and be able to compare the volume, tone, and lean across publishers and date ranges. In this sequence of images, we show the user interacting with the dashboard to focus on the `Presidential Horse Race' subtopic (D) after starting with the default all-category view (A) and then clicking on the `Politics' category (B), the `2024 Election' topic within that, and finally the horse race subtopic within it (D). At each level of interaction, the user is provided the same controls to color by Category, Lean, or Tone, to filter by publishers and article type, and to select a date range to focus on (\textbf{D1}). Hovering on a segment displays a tooltip explaining what it represents and presenting the count and proportion of articles in it.}
    \label{fig:depth}
\end{figure*}

The Coverage view forms the default landing page of our dashboard and presents an overview of the topic-level and subtopic-level news coverage since the beginning of our data collection (January 1st, 2024). Figure \ref{fig:coverage_bar} shows this view which takes the form of a stacked bar chart coupled with a number of controls that enable users to slice the data along different dimensions (\textbf{D1)} or filter it to focus on a specific publisher/topic/time frame \textbf{(D2)}. Each bar in this view represents the number of articles published by a particular news source on the selected topic(s) within a given time frame. The array of controls at the top gives users the ability to change the visualization parameters in various ways. The first two toggles (\ding{172} and \ding{173}) allow users to vary the type of visualization (stacked bar or grid view, see Figure \ref{fig:coverage_grid}) and toggle between different types of article labels (category, tone, or lean), respectively (\textbf{D1}). Figure \ref{fig:coverage_bar}B shows the lean and Figure \ref{fig:coverage_bar}C shows the tone labels for the same set of articles. 

Control \ding{174} allows users to select a particular category, topic, or subtopic to focus on and can be used to zoom into specific news articles (\textbf{D2}). This functionality is shown as an interactive sequence in Figure \ref{fig:depth} which begins with all news categories, then zooms into `Politics' to `2024 Election' to `Presidential Horse Race'. Similarly, control \ding{175} allows users to pick the article type (news report, news analysis, or opinion) and the final two drop-down menus let users select a time frame and subset of publishers to focus on. The \emph{Normalized} toggle (control \ding{178}) allows users to directly compare the proportion of coverage on a given topic between different publishers, as shown in Figure \ref{fig:coverage_bar}A. Disabling this option allows users to see the absolute number of articles published on a given topic (Figure \ref{fig:coverage_bar}B and \ref{fig:coverage_bar}C).

In this stacked bar chart view, we use five colors to represent five levels of political lean (`Democrat', `Neutral Leaning Democrat', `Neutral', `Neutral Leaning Republican', and `Republican') and tone (`Very Negative', `Negative', `Neutral', `Positive', `Very Positive'). The length of each color segment is proportional to the number of articles with that label, allowing for a comprehensive view of the label distribution. Users can hover on any segment to see the number of articles that fall within that label, as can be seen in Figures \ref{fig:coverage_bar}B and \ref{fig:coverage_bar}C.
 
Figure \ref{fig:coverage_grid} shows the grid view of the Coverage dashboard, accessible via control \ding{173}. Instead of using stacked bars, this view separates each segment into individual rows, with the bar size in each cell indicating the number of articles a news source has published on that topic. The view provides a broad overview of the news coverage and allows users to easily see at a glance what publishers prioritize which topics over others (\textbf{D3}). The publisher with the highest number of articles on a given topic is represented by the tallest bar in the row, with the exact article count displayed at the top. This enables users to compare and interpret the relative heights of other bars accordingly. Similar to the stacked bar view, users can also see the lean and tone versions of this grid as in Figure \ref{fig:coverage_grid}B and \ref{fig:coverage_grid}C. Here, the colors indicate the average political lean or tone of each publisher's articles within a given news topic or subtopic. The grid view's summary statistics provide a complementary perspective to the full distribution of labels shown in the stacked bar chart.

To enable \textit{deep} exploration of the data, users can select from controls \ding{174}, \ding{175}, and \ding{176} (see Figures \ref{fig:coverage_bar}A or \ref{fig:coverage_grid}A) to filter the full set of articles in our database to the topics/subtopics, publishers, and time frame they are interested in (\textbf{D2}). The sequence in Figure \ref{fig:depth} shows what interacting with control \ding{174} looks like. Clicking on `Politics' takes a user from the view in Figure \ref{fig:depth}A to the one shown in Figure \ref{fig:depth}B. Subsequently selecting the `2024 Election' topic within that category, and then selecting the Presidential Horse Race subtopic within that, takes the user through the views shown in subfigures \ref{fig:depth}C and \ref{fig:depth}D. At each level, users can choose to filter according to their desired time frame and subset of publishers, and also choose to color the data by category or tone instead of lean (\textbf{D1}).

\begin{figure*}
    \centering
    \includegraphics[width=0.9\textwidth]{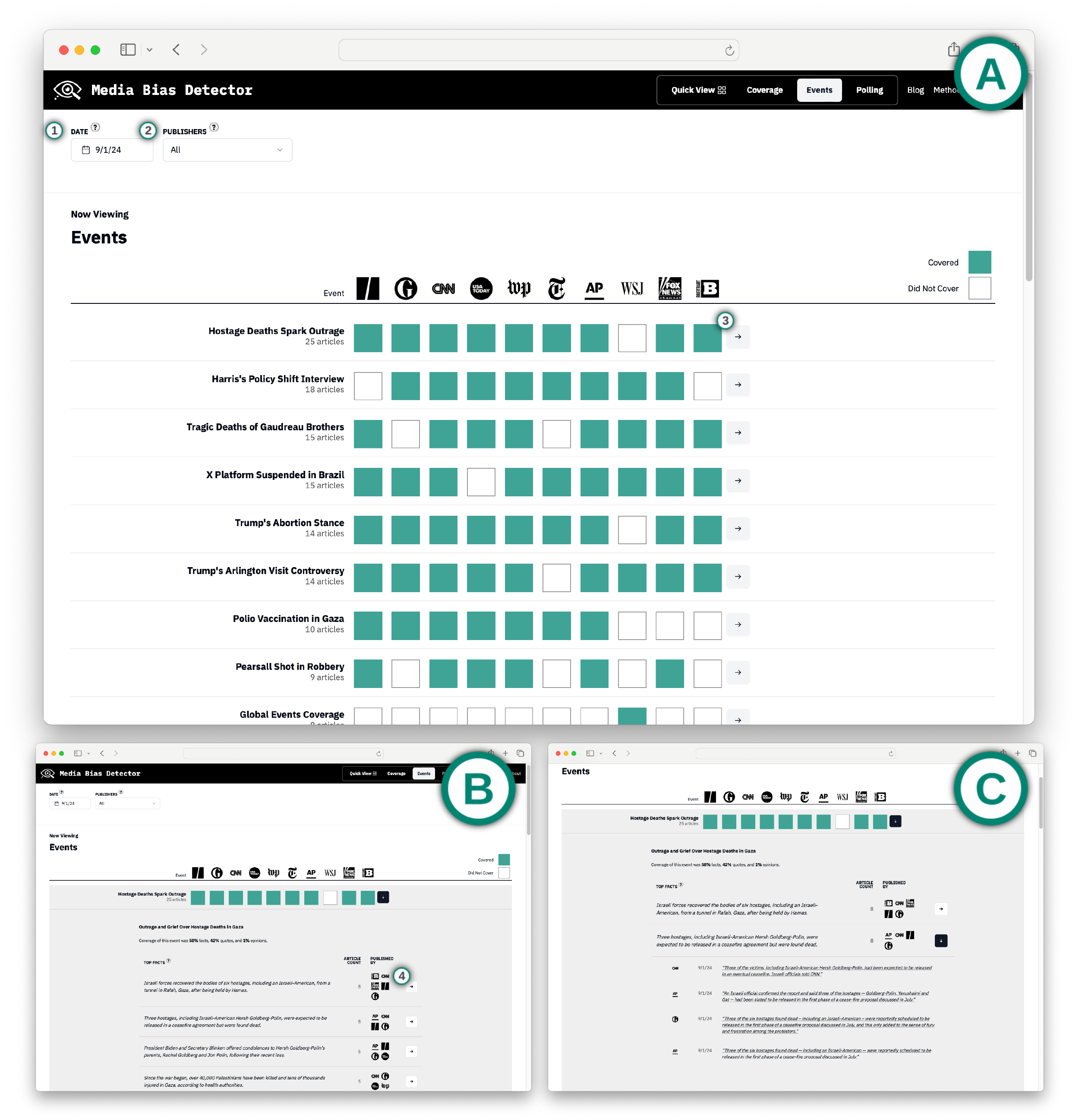}
    \caption{Complementary to the Coverage view, the Events dashboard shown here presents an event-level view of the news that focuses on the fast pace of the news cycle and caters to both \textit{broad} (\textbf{D1}) and \textit{deep} (\textbf{D2}) exploration of the news. It offers a quick overview of happenings from the past three days but allows users to dig deeper into each event and its major facts and compare selection and framing bias across publishers (\textbf{D3}). Each row is a news event and each cell represents whether it was covered or not by the respective publisher. The events are sorted by importance (measured as their amount of coverage) and we provide a summarized title for the event and a count of the number of articles about it across all publishers. Clicking on an event using the button on the right displays a detailed view (B) which contains the full event description as well as a summary of its sentence-level composition in terms of facts, quotes, and opinions. This view also lists the top facts about the event and shows which publishers mentioned or omitted certain statements. Clicking on any one of these top facts shows different variations of it (C) as they were written in the original news articles allowing users to compare their framing. Clicking on any one of these variations takes the user to the original article on the publisher's website.}
    \label{fig:events}
\end{figure*}

\subsection{Events Dashboard}
\label{sec:design_events}

The Events view of our dashboard is intended to capture the fast pace of the news cycle, providing users with an overview of major events and key information highlights. Figure \ref{fig:events} shows this view where rows represent different events, sorted in descending order of the number of articles written about them in the past day. If an event is covered by a particular publisher, its corresponding square will be colored turquoise; otherwise, it will be empty. At the top of the page, we provide users with controls to select the date range (control \ding{172}) and set of publishers to display (control \ding{173}).

To provide a more in-depth view of these events (\textbf{D2}), users can click on the arrow at the end of each row (control \ding{174}) to expand the event and see a more detailed description about it, as shown in Figure \ref{fig:events}B. These descriptions include a longer title which more accurately summarizes the event, as well as overall statistics on the types of sentences found in articles about it (facts, quotes, or opinions). This view also displays the top facts about the event that have been mentioned by different publishers in their coverage of it. As shown in Figure \ref{fig:events}C, clicking on the arrow next to a top fact (control \ding{175}) displays various phrasings of it from different articles, and clicking on any article listed takes the user directly to the story on the publisher's website. This functionality, grounded in comparative analysis (\textbf{D3}), allows a user to compare and contrast not only how much coverage (selection bias) has been given to an event but also how this coverage has been presented (framing bias).

\section{Evaluating the \mbd}
\label{sec:study1}

To understand how the \mbd\ could benefit experts who study news and media, we conducted interviews with participants from academia, industry, and newsrooms, all of whom work closely with news content, but approach it from different disciplines. In addition to need-finding, we also conducted a within-subject evaluation study where we compared using our tool to baseline media bias detection strategies. This evaluation helped us understand how the \mbd\ complements existing tools, and we used feedback gathered from this study to inform improvements for the tool's next iteration. 

\subsection{Participants}

\label{sec:study1_participants}

We recruited 13 participants (7 women, 6 men, aged 24-63).\footnote{Two participants, P10 and P12, were unable to complete the second task evaluation and post-task interview due to time constraints, and we were unable to reschedule their interviews prior to the submission. Their evaluation results are dropped from the quantitative analysis, but we still discuss insights from their interviews in our findings.} Potential candidates were identified through personal networks, including referrals from colleagues, friends, and word of mouth. To ensure unbiased evaluations, participants were selected specifically for their lack of prior exposure to the tool. This design decision allowed us to collect immediate feedback on the tool's ability to engage and effectively communicate its purpose during a user's first interaction. By focusing on first-time users, we were able to identity potential areas for improvement that might be less apparent to users more familiar with the tool. All participants had undergone rigorous training as either researchers or writers for at least three years, enabling them to provide well-thought-out perspectives on news media. Their areas of expertise spanned various domains, including communications, political science, and journalism, with differing levels of familiarity with media bias. Detailed information about the participants is provided in Table \ref{tab:participants}. 

\begin{table*}[h!]
    \centering
    \small 
    \setlength{\tabcolsep}{3.5pt} 
    \renewcommand{\arraystretch}{1.1} 
    \caption{Participant Characteristics in the User Study with Experts. \emph{Current Role} and \emph{Area of Expertise} are self-reported descriptions. \emph{Exp. (Yrs.)} refers to the participant's experience working in their domain. \emph{Bias Familiarity} refers to the participants' self-perceived media bias familiarity.}

    \begin{tabular}{cp{2cm}p{4cm}p{4cm}cc} 
    \hline
    \textbf{ID} & \textbf{Domain} & \textbf{Current Role} & \textbf{Area of Expertise} & \textbf{Exp. (Yrs.)} & \textbf{Bias Familiarity} \\ \hline
    1 & Communications & Pre-doctoral Research Assistant and PhD Student & Digital Political Communication & 3-5 & High \\
    2 & Communications & Pre-doctoral Research Assistant & Environmental Communication & 3-5 & High \\
    3 & Communications & PhD Student & Tech Innovation for Cultural Protection & 3-5 & High \\
    4 & Journalism & Writer & Political Journalism & 5-10 & High \\
    5 & Communications & PhD Candidate & Political Communication, Journalism, AI & 10+ & High \\
    6 & Political Science & PhD Candidate & Political Behavior, N. American Politics, Misinformation & 3-5 & Medium \\
    7 & Communications & PhD Candidate & Social Media, Health Communication & 3-5 & Medium \\
    8 & Journalism & Director of News \& Media & Editorial Leadership, Tech in Journalism & 10+ & High \\
    9 & Political Science & PhD Candidate & Racial and Ethnic Politics, Black Politics & 5-10 & Low \\
    10 & Journalism & Retired; Former Director of Content and Managing Editor & Digital News, Social Media, Leadership & 10+ & High \\
    11 & Journalism & CEO of Journalism/Content Consulting Company & Journalism and Content Strategy & 10+ & High \\
    12 & Journalism & Guideline Architect; News \& AI Strategist & AI \& Content Strategy, Policy Development & 10+ & High \\
    13 & Journalism & Media Reporter & Media Studies & 5-10 & High \\ \hline
    \end{tabular}
    \label{tab:participants}
\end{table*}

\subsection{Task Design}
\label{sec:study1_task_design}

Each participant engaged in two consecutive tasks, where the first task required the use of a baseline online search and the second used the \mbd. These tasks were designed to simulate a realistic scenario in which participants evaluate media bias across different news publishers, setting up a practical comparison between traditional methods and our tool. 

In the first task, participants were presented with a scenario where they imagined discussing with friends the news sources they followed for the 2024 election. Participants selected two publishers to analyze from a predetermined list of ten publishers, corresponding to the ten publishers on the dashboard, described in Section \ref{sec:design}. To allow for a more personalized experience, participants were given the flexibility to choose two publishers from the list that aligned with their own interests or questions they wanted to explore. This approach prevented potential biases that might arise from pre-selecting specific publishers, which might have influenced the types of insights participants could generate. It also allowed us to capture variations in themes, editorial choices, and general nuances across publishers. Participants were asked to explore how selection and framing biases manifested in these outlets and influenced their news coverage. They were given 10 minutes to evaluate the overall bias of the two newspapers using any available tools or platforms known for assessing media bias and prepare a brief oral summary comparing the two. For the second task, participants engaged in the same scenario, evaluating the same two newspapers, but this time using our tool. We provided participants with a URL to access the \mbd\ and gave them 10 minutes to assess the overall bias of the publishers  (see Appendix \ref{sec:apdx_task} for task scenarios). 

After each task, participants answered a set of structured Likert-scale questions designed to gather quantitative feedback on their experience, focusing on the affordances of the tools, how effectively they supported users in recognizing bias, making comparisons across publishers, and engaging with media content (outlined in Section \ref{sec:measures}). The evaluation after Task 2 also included Likert-scale questions aimed at capturing participants' reflections on the \mbd’s usability and intuitiveness (see Appendix \ref{appendix:survey_questions} for survey questions). This two-task design ensured consistency in the evaluation process while providing participants an opportunity to explore and assess the affordances of both an existing tool and our proposed tool for detecting media bias. 

\subsection{Quantitative Measures}
\label{sec:measures}

We used a mixed-methods approach to evaluate the \mbd. In addition to collecting qualitative feedback on how participants used our tool, we also measured key dimensions of the tool's impact: bias identification and awareness (RQ1), comparative analysis (RQ1), and user engagement (RQ3). To quantify these aspects, we asked participants to rate their agreement on a 7-point Likert-scale (1 = Strongly Disagree, 7 = Strongly Agree), similar to prior work \cite{think_fast, reza_abscribe}. Furthermore, to address RQ1, we assessed subjective workload using the NASA-TLX procedure with weighting \cite{hart2006nasa}, a method commonly used in HCI research to measure a user's cognitive demand when performing a task. 

\begin{enumerate}
    \item \textbf{Bias Identification and Awareness}
    \begin{itemize}
        \item \textbf{Selection Bias}: The news gives disproportionate attention to specific topics.
        \item \textbf{Framing Bias}: The way news stories are presented, including the choice of language, influences how people perceive events being reported.
        \item \textbf{Bias Awareness}: This method sharpened my awareness of bias in news content.
        \item \textbf{Critical Thinking}: This method encouraged me to think critically about the news articles I read.
        \item \textbf{Effectiveness}: This method was effective at helping me identify media bias.
    \end{itemize}
    
    \item \textbf{Comparative Analysis}
    \begin{itemize}
        \item \textbf{Qualitative Comparison}: This method offered useful descriptive information for comparing bias across different publishers.
        \item \textbf{Quantitative Comparison}: This method offered useful quantitative measures for comparing bias across different publishers.
    \end{itemize}

    \item \textbf{User Engagement and Application}
    \begin{itemize}
        \item \textbf{Proactive Sharing}: I am interested in sharing and discussing the information I've learned about media bias with my family and friends.
        \item \textbf{Practical Application}: I am interested in applying the information I've learned about media bias in my daily news consumption.
    \end{itemize}
\end{enumerate}

\subsection{Procedure}
\label{sec:study1_procedure}

The study sessions began after participants signed a consent form and completed a survey that collected demographic data, professional experience, and self-perceived familiarity with media bias. The study was conducted via one-on-one Zoom calls, with participants granting permission to record the session. Each interview lasted between 45 and 75 minutes. The session started with a brief introduction, during which the interviewer informed the participant that their feedback would be used to improve a tool designed to help news consumers navigate the vast amount of news content, particularly during the 2024 election cycle. 

During the pre-task interview, we explored participants' general media consumption habits, preferred news sources, and their engagement with news and politics. We also asked for their views on how news outlets may emphasize certain topics while overlooking others (selection bias) and how the tone and language of articles can influence readers' perceptions (framing bias). Next, participants undertook the sequential 10-minute media bias comparative analysis tasks. Participants were encouraged to think aloud as they performed both tasks \cite{ramey_2006, nielsen2002getting}. Before the \mbd\ task, we showed a brief walkthrough video highlighting its main features. At the end of each task, we provided participants with a survey link to complete the evaluation questions, including Likert-scale and NASA-TLX questions.

The session concluded with a semi-structured discussion about their experiences using the tool. The discussion covered topics such as the tool’s potential impact on the participant's work or research, its usefulness in the context of the 2024 election, and how it compared to other tools they had previously used. Participants were also asked to suggest additional features and provide feedback on improvements that could enhance their experience with the tool (see Appendix \ref{appendix:guiding_questions} for guiding questions).

\subsection{Analysis}
\label{sec:study1_data_analysis}

We collected interview transcripts, observation notes, and a set of quantitative responses for each task. We coded and analyzed the qualitative data using Miro \cite{miro}, which allowed us to collaboratively code the data using  digital sticky notes \cite{johnson_miro_2022,burgess2021sticky,geraci2014work}. Four team members manually reviewed the interview transcripts noting key points. From these notes, we used an inductive coding approach to generate initial descriptive codes and identify relationships between themes discussed in participant quotes \cite{braun2006using}. Subsequently, our team reviewed the codes and preliminary themes with the assistance of an AI summarization tool and had multiple group discussions to note similarities and differences before agreeing on the main broad themes we learned from our expert interviews \cite{mcdonald_2019}. 

\begin{figure}
    \centering
    \includegraphics[width=\linewidth]{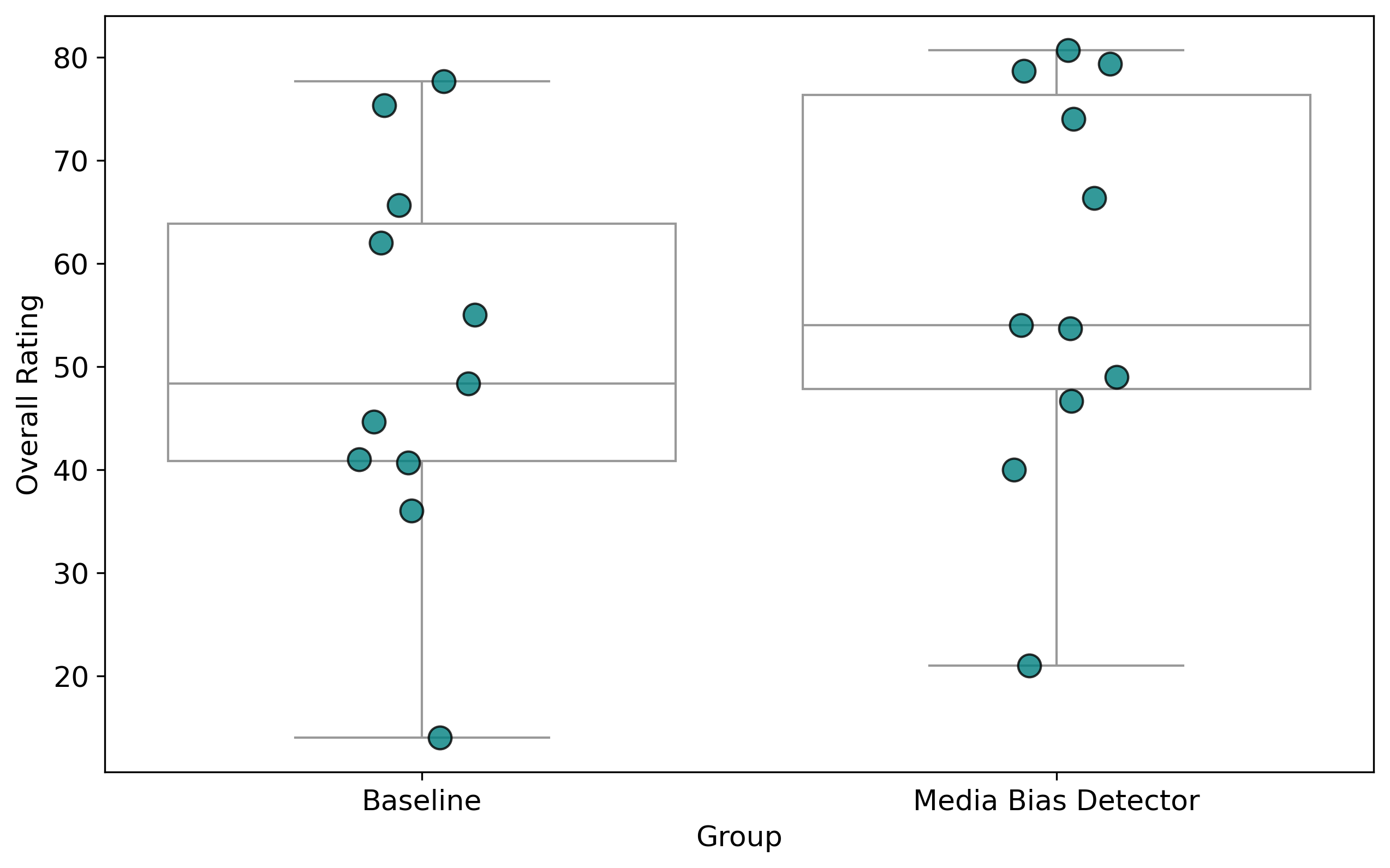}
    \caption{Overall results on the NASA-Task Load Index (NASA-TLX)'s measures of subjective workload for the Baseline and Media Bias Detector task evaluations in the user study with experts.}
    \label{fig:nasa}
\end{figure}

After analyzing the data, we reviewed the quantitative measures collected, including the NASA-TLX scores and Likert-scale responses. The boxplot in Figure \ref{fig:nasa} shows the overall results of the NASA-TLX ratings for subjective workload, comparing the baseline tool with the \mbd. While the \mbd\ received slightly higher overall ratings, indicating that participants found it somewhat more challenging to use, the difference is not substantial. The overlap between the distributions suggests that the \mbd’s task workload is comparable to the baseline tool, and several participants rated both tools similarly. These results are encouraging for the \mbd, as the tool is inherently more complex and offers more features than typical baseline tools, including those selected by participants. Despite this added complexity, participants did not find it overwhelmingly harder to use. This suggests that with continued improvement, such as providing more tutorials or onboarding, we can reduce the tool's perceived difficulty and help users take full advantage of its advanced features without significantly increasing the cognitive load.

\begin{figure}
    \centering
    \includegraphics[width=\linewidth]{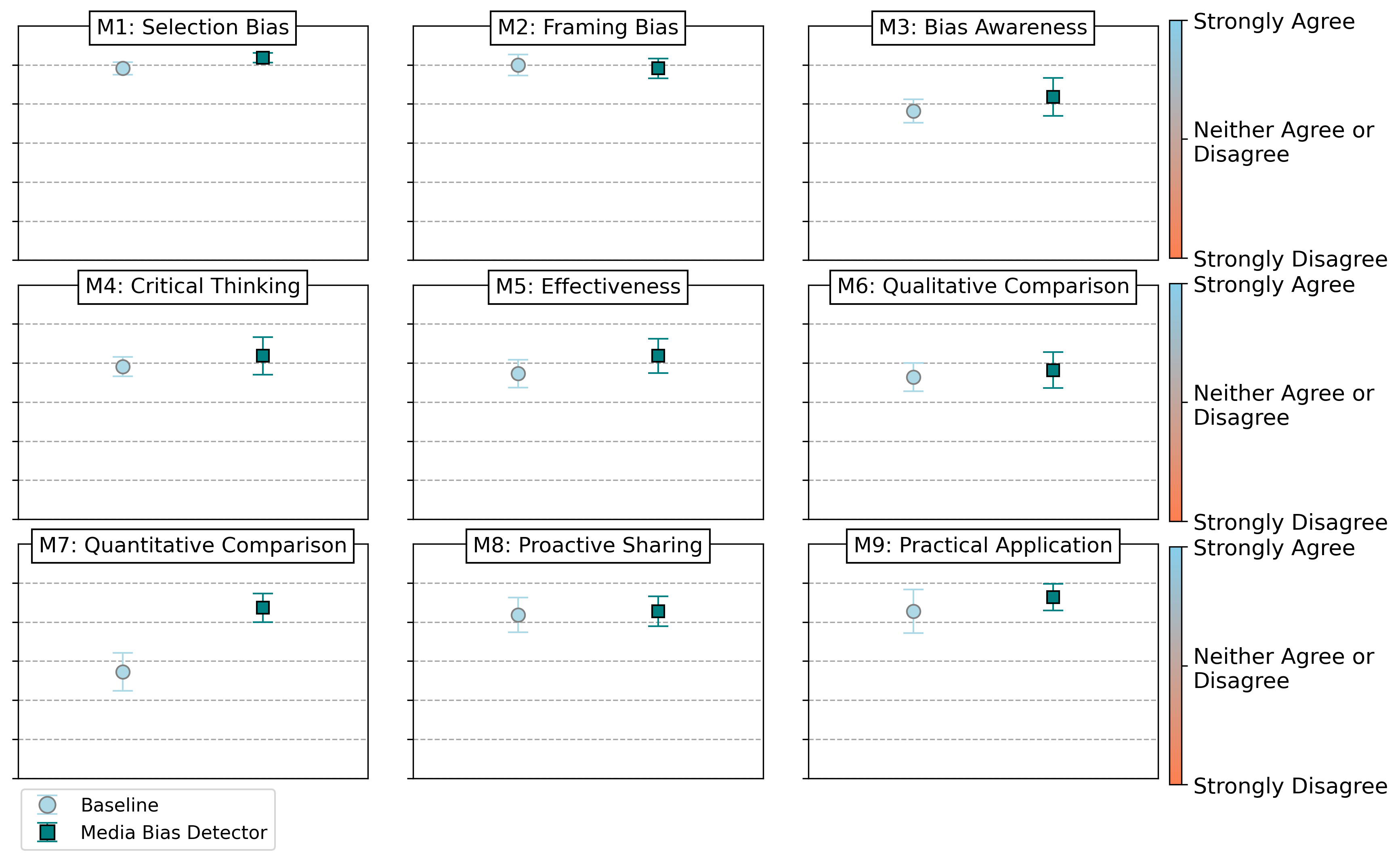}
    \caption{Average responses to Likert-scale measures on bias identification and awareness (M1-5), comparative analysis (M6, M7), and user engagement (M8, M9) in the user study with experts.}
    \label{fig:likert-agreement}
\end{figure}

The results of the Likert-scale responses in Figure \ref{fig:likert-agreement} show minimal variation between the baseline tool and the \mbd\ for most metrics, suggesting that short-term use of the tool may not lead to significant shifts in users' overall perceptions of media bias. The subplot describing \emph{M7: Quantitative Comparison} in Figure \ref{fig:likert-agreement} shows that only responses to "this method offered useful quantitative measures for comparing bias across different publishers" had a notable increase in the \mbd\ group. Given that media bias perceptions are often deeply ingrained and difficult to shift in a short time frame, these findings align with our expectations. Our small sample size also may have contributed to the lack of significant differences across most metrics. However, the results generally reinforce the idea that the true value of the \mbd\ may emerge over long-term use, as users engage more deeply with its features and gain more exposure to its capabilities. We explore this further in our follow-up survey (see Section \ref{sec:prolific_survey}) that focuses on other dimensions, including users' long-term beliefs about the tool’s potential impact, rather than just the measures evaluated in this initial study.

\section{Findings}
\label{sec:qual_findings}

The qualitative data from our interviews provided rich insights, addressing each of our main research questions while also uncovering additional findings. Common themes such as usability and educational potential offered a deeper understanding of the tool's strengths, contributing to RQ1. Other themes, such as trust in AI-driven tools and identifying the most suitable audience for the \mbd, aligned closely with RQ2 and RQ3, respectively. In this section, we explore the major themes that emerged from our expert interviews.

\subsection{RQ1: Functionality and Impact of the \mbd}
\label{sec:qual_findings_usability}

\subsubsection*{Interactive Presentation of Complex Information}
\label{sec:qual_findings_visual_presentation}

Overall, participants were impressed that the \mbd\ transformed a complex dataset into an easy-to-navigate interface. In particular, the tool's emphasis on progressive disclosure allowed participants to engage with the content at varying levels of depth (\textbf{D2}). P4 expressed that they \emph{"appreciate the effort to make the subtopics super granular"} after diving deeper from the \emph{Politics} category, to the \emph{2024 Election} topic, to the \emph{Presidential Horse Race} subtopic on the Coverage dashboard. Others expressed that the visualizations effectively communicated information when needed, making it easier to focus on coverage they were interested in. P3 especially liked how \emph{"the tool can be as interactive as you need it to be"}. 

Regarding \textbf{D1}, multiple participants mentioned the value of being able to examine the data from both the lean and tone perspectives. While traditional media bias tools focus on average political lean scores, P3 highlighted the unique distinction between lean and tone, emphasizing the importance of going beyond political lean. P8 expanded on this, pointing out that tone is a less talked about form of media bias: \emph{"[The bias is] not about negativity, but rather the lack of solutions journalism. For decades, people have wanted more context"}. Furthermore, P4 stated outright that the \mbd's ability to group data by both tone and lean offers a more comprehensive way of assessing media bias: \emph{"It's always hard to know on what basis [people building those tool] are making those decisions... This is a huge improvement because it allows you to select different forms of evaluation"}, contrasting it with tools like Allsides, which typically use charts to map news outlets' political lean from left to right.  

Participants also highlighted opportunities for improvement. P4, for instance, valued the range of features but remarked, \emph{"there was a lot of information in front of me, and that was overwhelming"}. Others saw potential in enhancing the user experience by introducing a search bar feature which allows users to more precisely search for specific topics by keyword (P1, P6). P6 suggested that this feature could reduce some users' cognitive burden by eliminating the need to actively scan through topic and subtopic options.

\subsubsection*{Interactions Shaping Media Bias Perceptions}

Being able to customize the data not only personalized the experience with the tool but also influenced how participants perceived bias in news coverage. For example, P3 described how the process of focusing on particular topics and within specific date ranges helped them challenge their preconceived notions of some publishers' biases: \emph{"I come in [with] really strong opinions on what the biases are for each of these outlets... because I got to choose the topics and choose the time frame and choose looking across the lean and tone for different topics, ... It helped me to ground my thinking in real world evidence"}. Similarly, P9 praised \textbf{D3}, noting that even without explicit bias ratings, the Events dashboard allowed them to easily compare different news sources and independently analyze how they framed the same events: \emph{"It's nice to see how they're covering them differently in one place... It's still helping me assess bias because it's putting everything in a place where I can look at it and seek out broader coverage, look, and assess it for myself."}

While some participants mentioned ways the \mbd\ could help challenge and expand their current understanding of media bias, not all participants described a significant shift in their perceptions. Instead, many participants shifted the discussion towards how the tool might help others learn about media bias, particularly those less experienced than themselves.

\subsection{RQ1 and RQ3: Potential Value in Education and Research}
\label{sec:qual_findings_edu_value}

\subsubsection*{RQ1: Applications in Education}
\label{sec:qual_findings_edu_applications}

Several participants, particularly those studying journalism and communications, highlighted the tool's potential as a resource in media literacy classrooms (P1, P2, P3, P7). P3 expressed that the \mbd\ would be very useful for media literacy classes, whether at the college level or in more public-facing settings. Furthermore, P1 and P2, both communications students, expressed interest in sharing the tool with journalism professors at their respective institutions. Beyond the classroom, participants also highlighted the \mbd's ability to educate users to recognize blind spots in news coverage. P2 emphasized that the tool provided a straightforward way to demonstrate that one news organization covers an issue more or less than another. On the Events dashboard, P9 highlighted an example of what an everyday user could learn from the \mbd; during the Gaza ceasefire negotiations, only seven major articles were published, while most coverage focused on the Democratic National Convention. P9 found this quantitative measure useful in showing how a significant event was being overshadowed, stating that \emph{"Americans are kind of distracted by the DNC and aren’t able to pay attention to something that's really important, and that is reflected quantitatively here"}.\footnote{P9 was referring to the Events page on Wednesday, August 21, where 33 articles on the top news event were related to the 2024 DNC, while only 7 articles were related to the Gaza Ceasefire Negotiations.}

In addition to recognizing the tool's potential for media literacy education, some participants noted that its full benefit depends on users having a baseline understanding of media bias. P7 suggested the tool would be most impactful when paired with education on bias, while P13 emphasized that users would have a better understanding of the \mbd\ if the experience could start with their own personal beliefs. More specifically, P13 recommended integrating a feature that allows users to reflect on how they currently view a topic, and then explore the news organizations that are covering it in a way that fits their own understanding. These participants emphasized a key point: \emph{"education about what media bias is and how it can affect them actually is the most important step for people to actually use these tools, because if they don't know what that does to them, they wouldn't really care"} (P7). For everyday users, simply providing access to the \mbd\ may not be enough; they need foundational knowledge to fully engage with its features, which we reaffirmed in our follow-up survey (Section \ref{sec:prolific_survey}).

\subsubsection*{RQ3: Applications in Research}
\label{sec:qual_findings_supporting_research}

The \mbd\ was also recognized as a valuable tool for research, particularly for supporting mixed-methods communications research. P1 and P2 highlighted its strength in providing quantitative comparisons that can help set the stage for more in-depth qualitative analysis \textbf{(D3)}. P2 noted that the \mbd\ offered a straightforward solution for obtaining quantitative evidence, saving time compared to relying on specialized organizations like Media Matters or searching through existing studies for relevant data. \emph{"[Need] to justify a news analysis?"} P2 declared, \emph{"here's results from a tool that justify it"}. In political science research, P6 and P9 proposed that the \mbd\ could be used as an experimental stimulus in studies on political behavior. P9 envisioned using the tool as an intervention in behavioral experiments that study how media exposure influences political attitudes, such as assessing \emph{"how [the tool's information] shapes your view of the American government's seriousness about ending the war in Gaza"}. Finally, to conduct high-quality research, P2 and P5 also pushed for a CSV download feature.

\subsection{RQ2: Challenges Surrounding Trust and Transparency in LLM-Driven Bias Detection}
\label{sec:qual_findings_ai_skepticism}

Participants recognized the \mbd's potential for shaping education and research but also expressed reservations about fully trusting the LLM-based classifications. For some, ensuring that humans were reviewing the data was the most important factor in building trust. P7, for example, immediately asked if human annotators were cross-checking the LLM's classifications and emphasized that it was important for them to know that real people were involved in the process. 

Their skepticism was rooted in doubts about the LLM's ability to capture the nuances of media bias, particularly in sensitive topics, such as child tax credits or international conflicts (P1, P9). P5 pointed out that while they felt comfortable believing that LLMs can easily automate formulaic news stories like financial reports, they needed more evidence to believe that the models can properly classify subjective news content. Similarly, P9 wanted explanations for or article examples of the LLM's labels for "Strong Democrat" or "Strong Republican" on topics with varying levels of polarization, noting that political distinctions may be harder to define for more nuanced issues. 

Several participants suggested that adding features to increase transparency could help build trust (P1, P2, P9). P1 expressed, \emph{"For me, someone who is a bit more skeptical of AI research in general, being able to see the articles [labeled under a topic] would make me feel better"}. P2 also suggested that having access to article texts would help them evaluate how the LLM processes articles with mixed tones, particularly in climate change coverage, where optimistic, positive outlooks may exist within negative stories \cite{ojala2021anxiety}. This feature would allow them to determine if the LLM captures subtle shifts in tone or oversimplifies it as neutral. 

Nevertheless, many participants noted that spending more time reviewing the Methodology section might help alleviate their concerns (P2, P6, P7, P9, P11, P13). Others saw LLM-driven bias detection not as a concern but as a practical solution for scalability. P11 shared the desire to understand the \mbd's inner workings but recognized the importance of AI in handling large-scale, dynamic data. During the task, they repeatedly "calibrated" themselves, using their prior knowledge to compare the tool's lean and tone ratings across the political spectrum to feel more confident in the \mbd's trustworthiness.

Overall, increased visibility into the \mbd's decision-making process could mitigate the skepticism expressed by participants. Comments from multiple participants suggested that their trust in the tool was closely linked to their own ability to see, interpret, and agree with the AI model's decisions. Thus, providing more transparency about the article texts, human-in-the-loop's weekly analysis, and the LLM's classification process could help users feel more comfortable using the \mbd.

\subsection{RQ3: Who is the User? Clarifying the \mbd's Audience}
\label{sec:who_is_user}

A recurring theme among some participants (P8, P10, P11, P13) was an encouraging attitude toward the \mbd's development. However, many did not see themselves as the primary users. P10 mentioned that they had already developed methods for evaluating bias independently because their role at a news aggregator had required assessing media bias and deciding what content to publish. Meanwhile, P8 provided a reflective observation, remarking, \emph{"\textit{Can} you teach media bias [sic] about media bias? Because a lot of media people don’t necessarily think that they are biased in certain ways."} This highlights the challenge of introducing media bias tools to professionals who, confident in their expertise, may be less inclined to fully engage with a tool designed to reveal what they feel they already know.

For other experts, there was no clear consensus on the tool's target audience. Some, like P3, found the tool useful for encouraging self-reflection, noting that it helped them \emph{"think through my own thoughts--what I thought were the biases versus what actually are the biases"}. Others, like P5, believed the tool was well-suited for researchers with a nuanced understanding of media studies but felt it might be too detailed for regular news consumers. In contrast, P1 suggested that the tool could resonate with everyday people who are seeking balanced news, especially during the 2024 election.

\section{Follow-Up Survey: Evaluating the Broader Population}
\label{sec:prolific_survey}

Through our 13 semi-structured interviews, we gained valuable feedback on the \mbd’s usability, its educational and analytical value, and trust concerns associated with LLM-driven media bias detection. It also became clear that while some participants saw value in the tool for their own use, many perceived it as being more relevant for other users rather than themselves. In particular, the feedback suggested that understanding the tool’s impact on everyday news consumers would be crucial for further refining its design and assessing its broader applicability. To address this, we conducted a follow-up survey targeting everyday news consumers. 

\subsection{Procedure}

We recruited 150 participants (51\% male, 46\% female, 3\% other\footnote{The "other" category comprises two participants who preferred not to disclose their gender and two participants who self-identified as non-binary. This categorization does not significantly impact the broader analysis, as gender is not a primary parameter in our study.}) through the crowdsourcing platform, Prolific, with the sample stratified by political party to balance Democrats, Republicans, and Independents (see Appendix \ref{sec:app_study2}, Figure \ref{apx:population}). Participants first completed a demographic survey, followed by questions about their media consumption habits, familiarity with media bias, and prior use of bias detection tools. The main survey was divided into three stages: pre-tool exposure, training and exploration, and post-tool exposure.

In the pre-tool exposure stage, participants answered general questions about their perceptions of selection and framing bias in the media, as well as a set of more targeted questions related to specific issues covered in the upcoming training task. Participants were asked to give their best guesses to the following questions. The questions were selected to illuminate selection biases in coverage by various news outlets on important and timely topics during the 2024 election campaign. We chose questions that might challenge people's existing misconceptions about media bias, so that observing the actual coverage on the dashboard would better capture their interest.

\begin{enumerate}
    \item Which news categories tend to get mostly negative coverage?
    \item Which media outlets covered the topic of Biden's age more frequently?
    \item What is the political lean of The Wall Street Journal's economic coverage?
    \item What is the most prominent event covered by the media this past week?
\end{enumerate}

\begin{figure}[htbp]
    \centering
    \begin{subfigure}{0.45\textwidth}
        \centering
        \includegraphics[width=\linewidth]{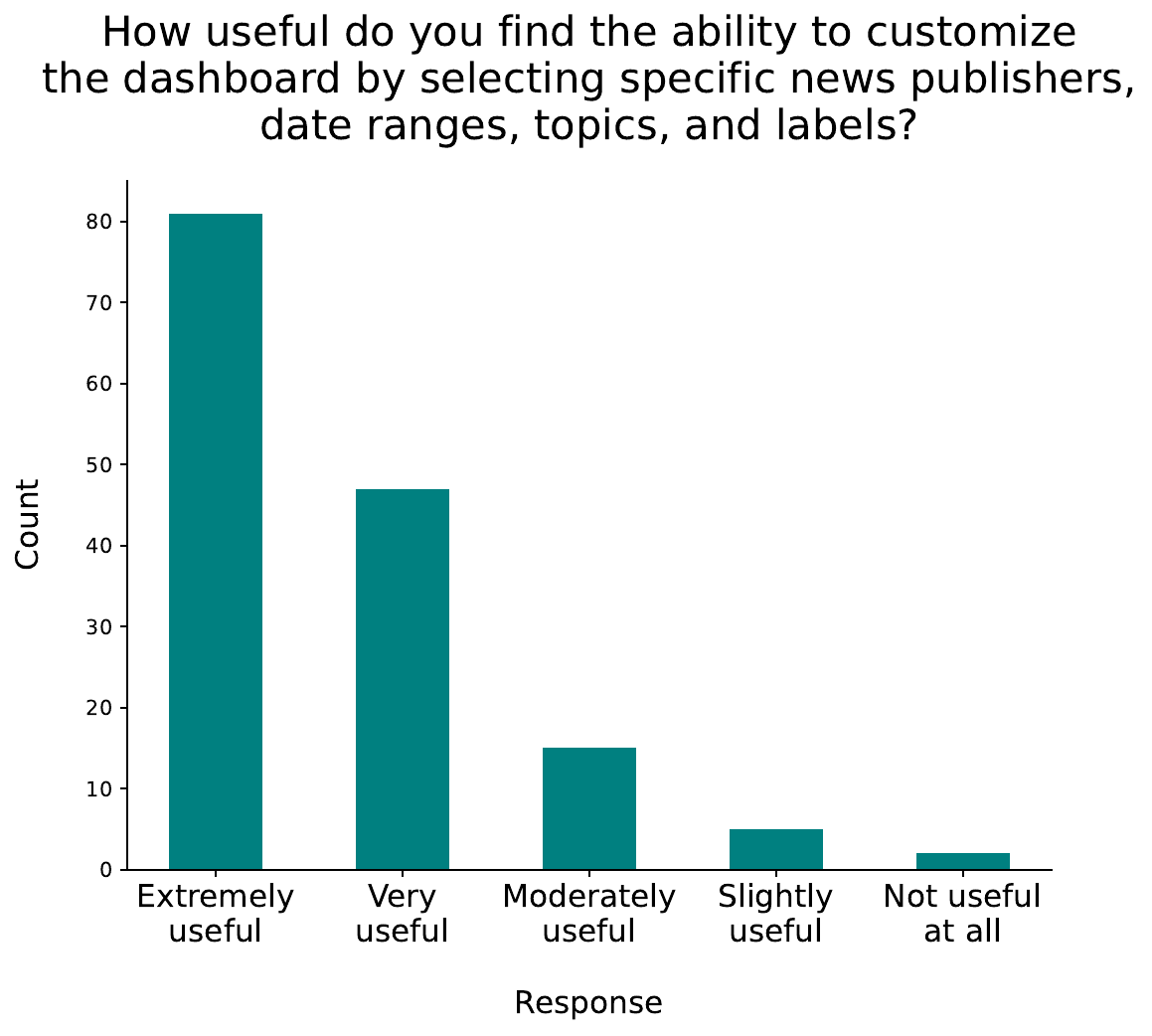}
        \caption{Customization usefulness.}
        \label{fig:design-subfig1}
    \end{subfigure}
    \hfill
    \begin{subfigure}{0.45\textwidth}
        \centering
        \includegraphics[width=\linewidth]{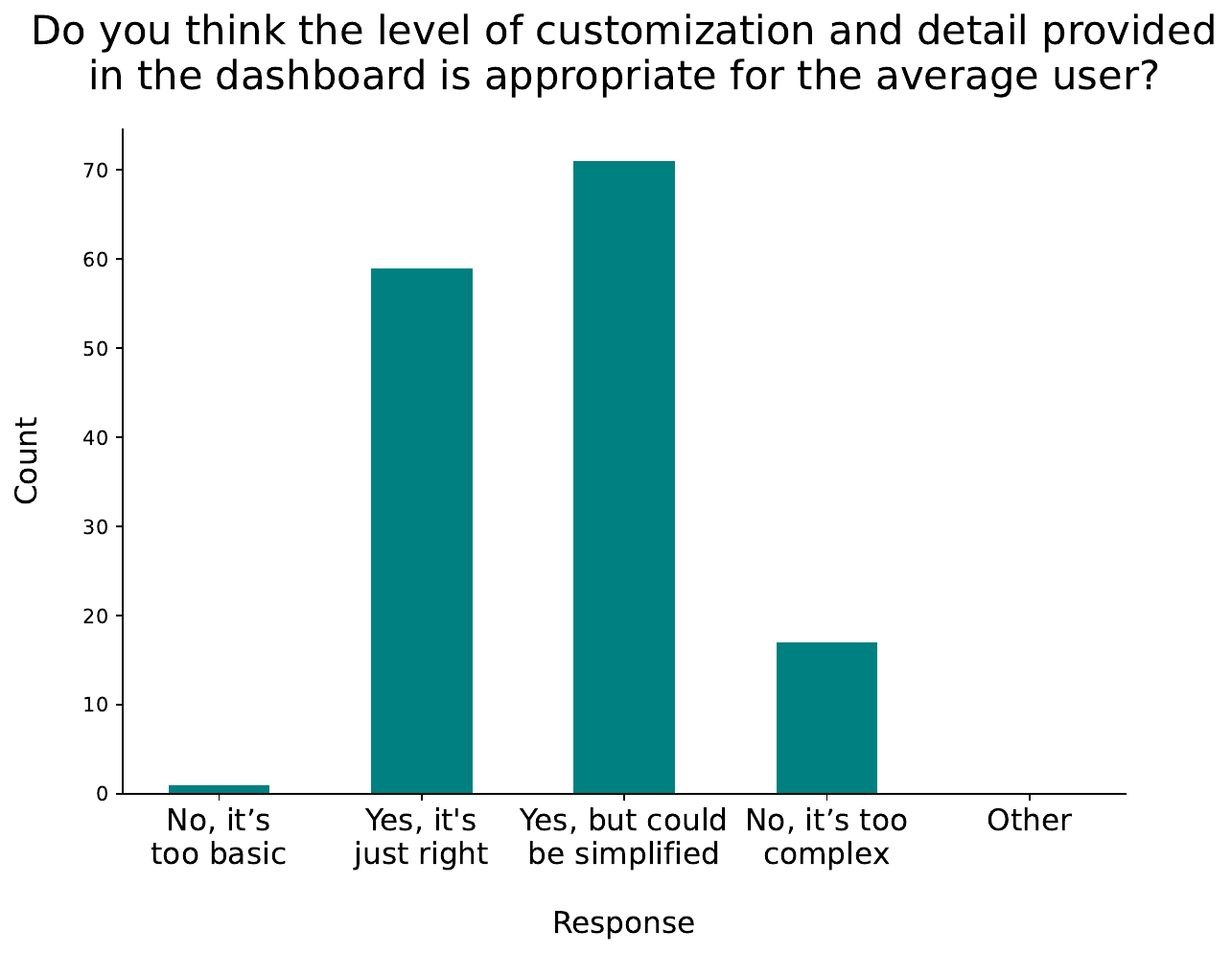}
        \caption{Customization appropriateness.}
        \label{fig:design-ux2}
    \end{subfigure}
    \caption{In response to questions about the \mbd's customization features, the vast majority of participants stated that they found these options to be `extremely' or `very' useful. However, when asked whether this level of customization would be appropriate for an average user, most recommended that it could be simplified, suggesting a steep learning curve for the tool's current design.}
    \label{fig:ux_prolific}
\end{figure}

During the training and exploration stage, participants first watched a brief walk-through video of the \mbd. Afterward, they followed guidelines to explore the tool and answer the same questions asked in the pre-tool exposure stage. This part of the task design aimed to teach participants how to use different features of the tool. Following this guided experience, participants were given five minutes to use the \mbd\ to explore media bias related to a topic of their choice and report their most interesting findings. This exploration was intended to encourage deep engagement with the tool, giving participants the freedom to find patterns in the data and consider how they might use it on their own.

Finally, in the post-tool exposure stage, participants revisited the specific questions asked in the pre-tool exposure stage to measure whether the training task had any short-term impact on their responses. However, recognizing that a brief five-minute interaction is unlikely to change deeply held views about bias in the news, as discussed in Section \ref{sec:study1_data_analysis}, we also asked them directly whether they believe that using the \mbd\ over time could change their view on media bias. Additionally, participants provided feedback on the tool's complexity, customization options, and their trust in AI-driven classifications (see Appendix \ref{sec:app_study2} for survey questions).

\subsection{Analysis}

\subsubsection*{User experience}

One of the primary features of the \mbd\ is the ability to view the data from many different perspectives (\textbf{D1}). To understand whether ordinary users like this feature, we asked them if they found this ability to customize the visualizations useful, and whether they thought that this level of customization was appropriate for other ordinary users. Figure \ref{fig:ux_prolific} shows their responses. A little over half the participants (81/150) said that they found the provided customization `extremely useful', while an additional one-third (47/150) said that they found it `very useful'. Interestingly, when asked whether they thought this level of customization and detail would be appropriate for the average user, around 40\% of respondents (59/150) said it was just right, but a slightly larger 47\% of respondents (71/150) said that it could be simplified.

We also asked for respondents' feedback on any features they found unclear or challenging to use, as well as suggestions for improvements. This input further mirrored the variation between 'just right' and 'simplified' seen in Figure \ref{fig:design-ux2}.  Some found it easy, with one saying, \emph{"the truth is I think it is very easy to use and learn,"} while others felt overwhelmed, noting, \emph{"there is so much to take in just in this one setting"} and \emph{"Still absorbing what is here"}. Participants also provided a wide range of ideas, suggesting improvements to the UI (\emph{"When I hover over different clickable options, it'd be nice to be reminded what each option means"}), data visualizations (\emph{"average person will probably want something more visual... Perhaps a wheel or pie chart structure"}), and usability (\emph{"I only want to do 2 or 3 clicks to get my information}"). 

\subsubsection*{Impact of using the \mbd}

\begin{figure}[htbp]
    \centering
    \begin{subfigure}{0.49\textwidth}
        \centering
        \includegraphics[width=\linewidth]{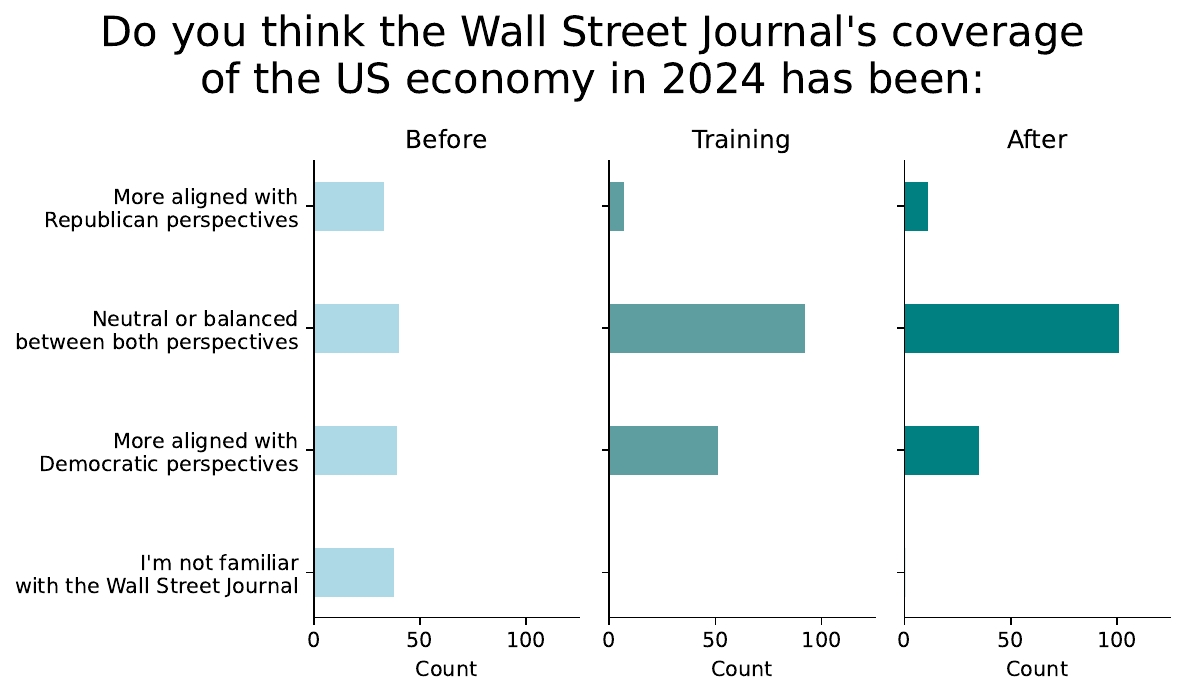}
        \caption{WSJ and the economy.}
        \label{fig:tone-subfig1}
    \end{subfigure}


    \begin{subfigure}{0.49\textwidth}
        \centering
        \includegraphics[width=\linewidth]{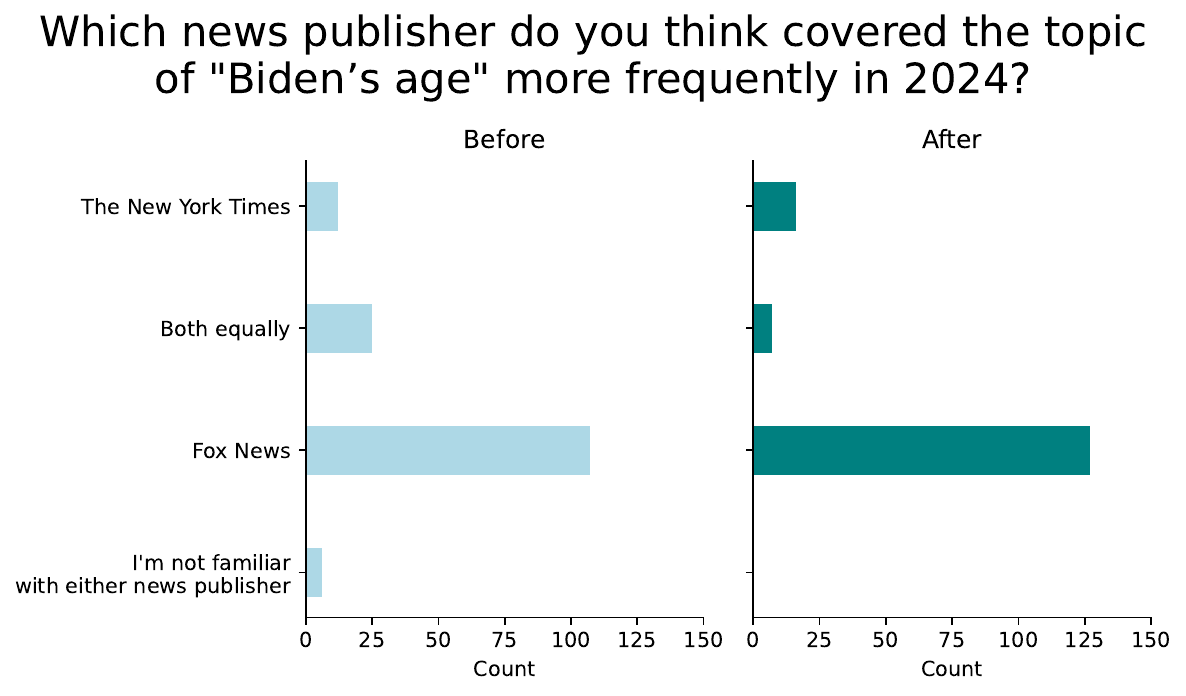}
        \caption{Biden's age: Fox vs. NYT}
        \label{fig:tone-subfig2}
    \end{subfigure}
    
    
    \begin{subfigure}{0.49\textwidth}
        \centering
        \includegraphics[width=\linewidth]{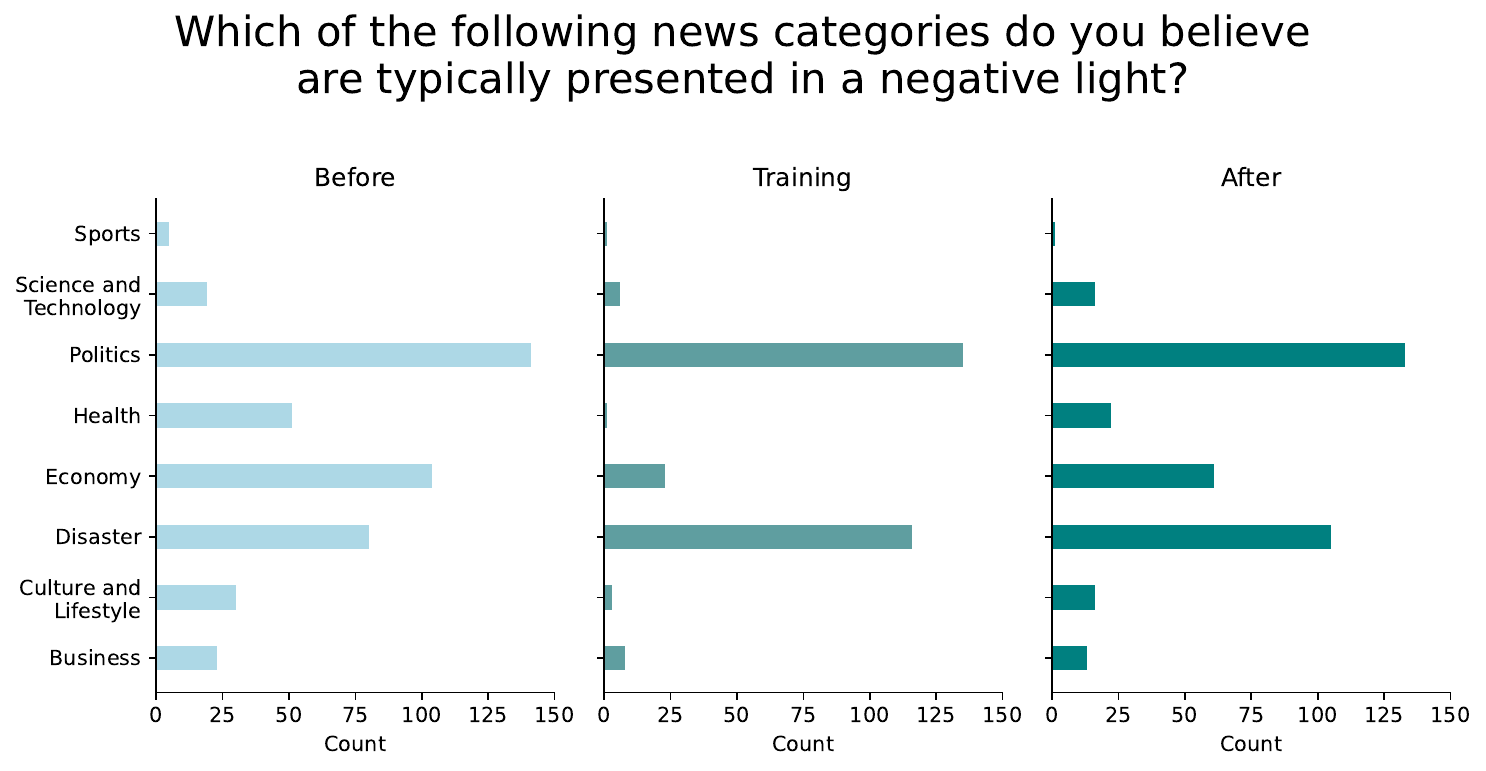}
        \caption{Negativity in the news.}
        \label{fig:subfig3}
    \end{subfigure}
    \caption{Participants showed a significant shift in their responses after using the tool, either correcting their priors where the dashboard's results showed otherwise, or strengthening them when the results matched their predictions.}
    \label{fig:tool_impact}
\end{figure}

We also investigated whether and to what extent the \mbd\ can change a user's perceptions of media bias. To do this, we asked participants about their views on news coverage for certain topics in the pre-tool exposure stage, had them answer the same questions during the training stage, and then asked again in the post-tool exposure stage. Figure \ref{fig:tool_impact} shows the impact using the \mbd\ had on their responses. In Figure \ref{fig:tool_impact}a, we asked users what they imagined the political leaning of the Wall Street Journal's coverage of the economy looked like in 2024. We received a fairly even response across the board: while a quarter of the respondents (38/150) were not familiar with the publisher, the rest were almost evenly split between pro-Democrat, pro-Republican, and neutral coverage. The training phase revealed that the \mbd\ classified a majority of the Wall Street Journal's economic coverage as neutral, and most respondents arrived at the same conclusion after using the tool. Furthermore, this had a significant impact on their post-training responses, with over two thirds of respondents (101/150) correctly stating that the publisher's coverage was indeed neutral. Some participants noted their surprise at learning this in a free-response question later in the survey: \emph{"I find it interesting that the Wall Street Journal really does seem to have the most balanced takes on business matters"}.

In some cases, however, the data found on the dashboard was less surprising and mostly confirmed people's pre-existing notions. Another question we asked survey participants was who they think covered the topic of President Joe Biden's age more: the New York Times or Fox News (\textbf{D3}). Figure \ref{fig:tool_impact}b shows the result: 71\% of respondents (107/150) answered Fox News, the correct answer, even before using the tool. During the training stage, participants were asked to provide the exact number of articles on Biden's age published by each of the two publishers, and 90\% of the participants (135/150) provided the correct answer using the tool. When asked the same question in the final stage of the survey, some participants adjusted their responses accordingly, resulting in 85\% of them (127/150) providing the correct answer in the end.

Finally, using the \mbd\ can help people tune their priors to be more in line with observed data. We asked participants which topics they thought were generally presented in a negative light in the news, where they could choose multiple topics at once. As shown in Figure \ref{fig:tool_impact}c, most participants correctly chose topics such as Politics (141/150) and Disaster (80/150)\footnote{Note that participants could select multiple topics.}, but many also selected Health (51/150), Economy (104/150), and Culture and Lifestyle (30/150), which are generally more neutral or positive. After using the tool, they encountered labels indicating that only Politics and Disaster had an overall negative tone across publishers, while Economy was covered negatively by some of them. Post-training, participants updated their answers accordingly, resulting in an increased number of votes for Disaster (105/150) and decreased number for Economy (61/150), Health (22/150) and Culture and Lifestyle (16/150).


\subsubsection*{AI Skepticism}

\begin{figure}[htbp]
    \centering
    \begin{subfigure}{0.45\textwidth}
        \centering
        \includegraphics[width=\linewidth]{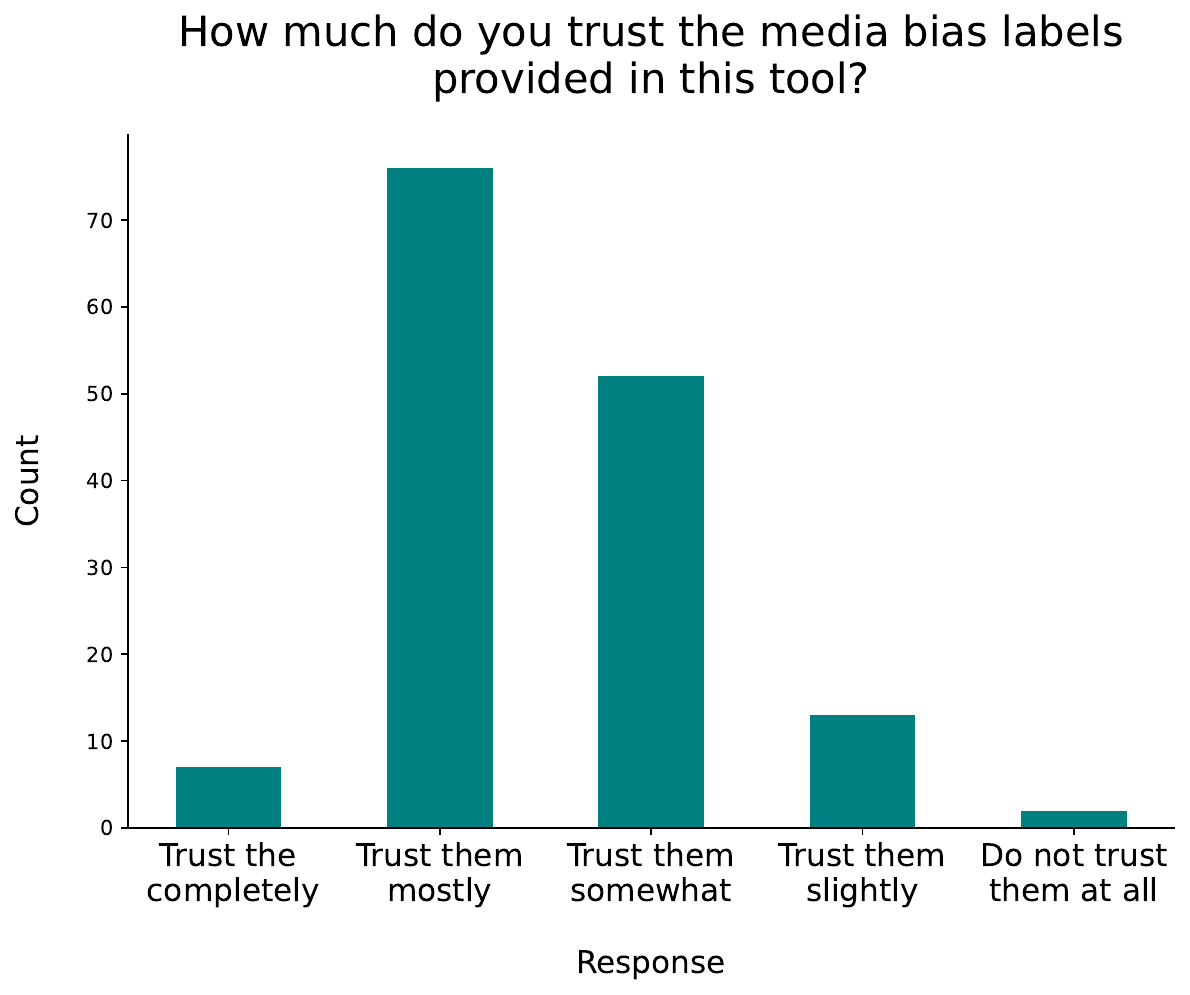}
        \caption{Trust in the  labels provided by the \mbd.}
        \label{fig:ai-subfig1}
    \end{subfigure}
    \hfill
    \begin{subfigure}{0.45\textwidth}
        \centering
        \includegraphics[width=\linewidth]{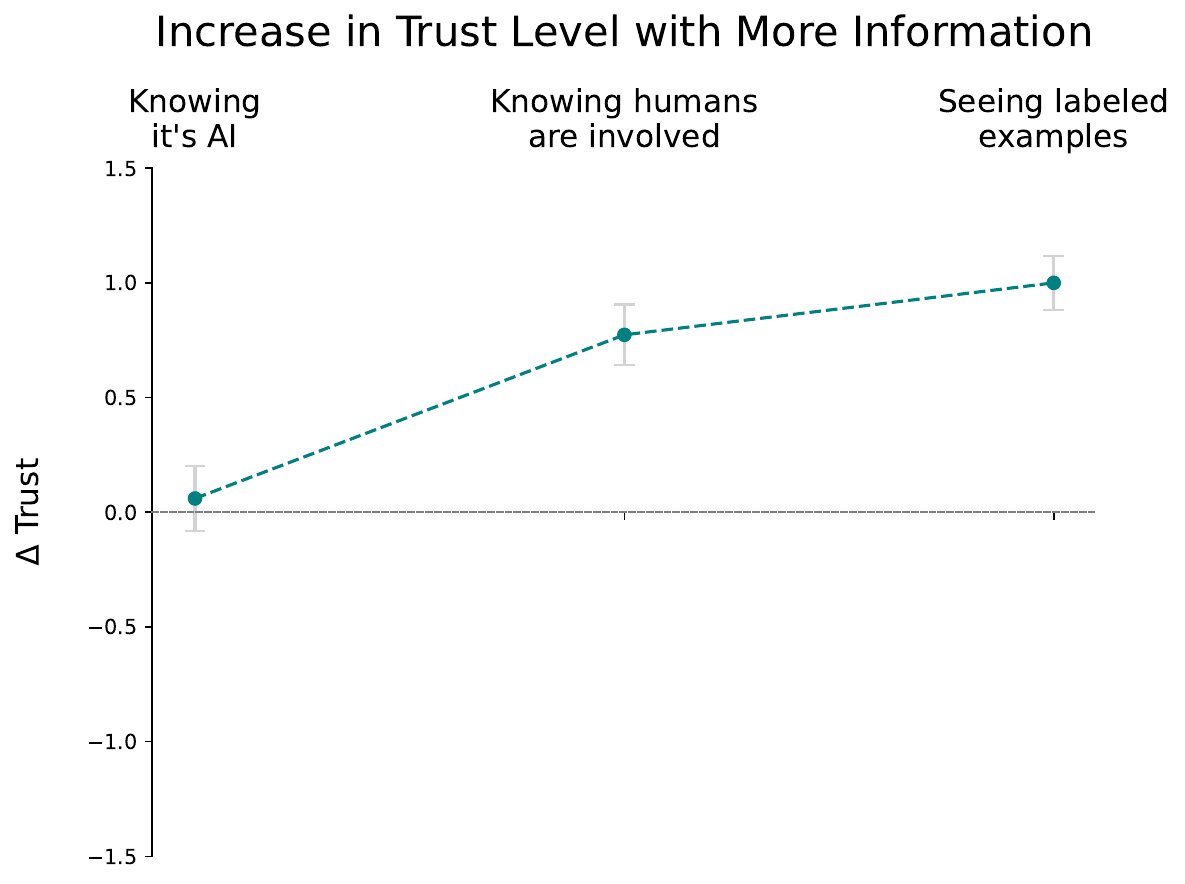}
        \caption{Change in level of trust with more knowledge of the tool's inner workings.}
        \label{fig:ai-subfig2}
    \end{subfigure}
    \caption{In contrast to our expert participants, the majority of our general public respondents expressed a significant amount of trust in the \mbd's labels. More interestingly, however, this trust did not waver after learning that the tool was powered by AI, and telling people about the involvement of humans in its labeling process increased this trust, as did the possibility of seeing labeled examples of articles so they could gauge it's quality for themselves. Point estimates show the average of the responses on a -2 to +2 discrete scale, while the error bars depict the associated 95\% confidence interval.}
    \label{fig:ai}
\end{figure}

In the user study with experts, many participants voiced skepticism about the underlying methodology and expressed a desire for more details, along with examples of how the labels were chosen (see Section \ref{sec:qual_findings_ai_skepticism}). To understand whether the general public exhibited a similar mistrust of AI, we asked them whether they trusted the \mbd's labels before knowing it was powered by AI, after knowing it was powered by AI, and after being told that the labels had been validated by human annotators. Figure \ref{fig:ai} shows how participants' responses varied. We found that most users exhibited a high degree of trust in the tool, with nearly half (76/150) stating that they trusted the labels `mostly', and another third (52/150) stating that they trusted them `somewhat'. Being told that these labels were generated by LLMs did not move the respondents' trust on average. Figure \ref{fig:ai}b shows the average of the responses to this question on a -2 to +2 discrete scale, while the error bars depict the associated 95\% confidence interval. We did not observe any significant variation in these trends by age group, suggesting that old and young people are equally (un)skeptical about AI. However, trust in the tool rose after respondents were told that humans are involved in monitoring data samples to ensure the LLMs are performing consistently. Participants also indicated that seeing examples of specific articles rated by the models would increase their trust in the labels even further. 

We also asked participants what additional information or features would increase their trust in the tool and received useful feedback. Some respondents wanted more details on the underlying methods (\textit{``An overview of how the AI system makes determinations would help me understand what's happening, which is likely to increase my trust in it."}) while others wanted to dig into the data and be able to \textit{``drill down to the level of headlines or blurbs"} and \textit{``take the time to look at specific examples and compare my judgements with the LLM"}. One of the most frequent pieces of feedback we received was a request for insight into the political leanings of the team that built the tool, with one participant asking for the tool to be \textit{``a collaboration of all diversities and political preferences"} and another saying their trust would increase if \textit{``independent organizations regularly audit the tool for accuracy and allow users to compare bias findings with other tools for added validation"}.

\subsubsection*{Target audience}

\begin{figure}[htbp]
    \centering
    \begin{subfigure}{0.45\textwidth}
        \centering
        \includegraphics[width=\linewidth]{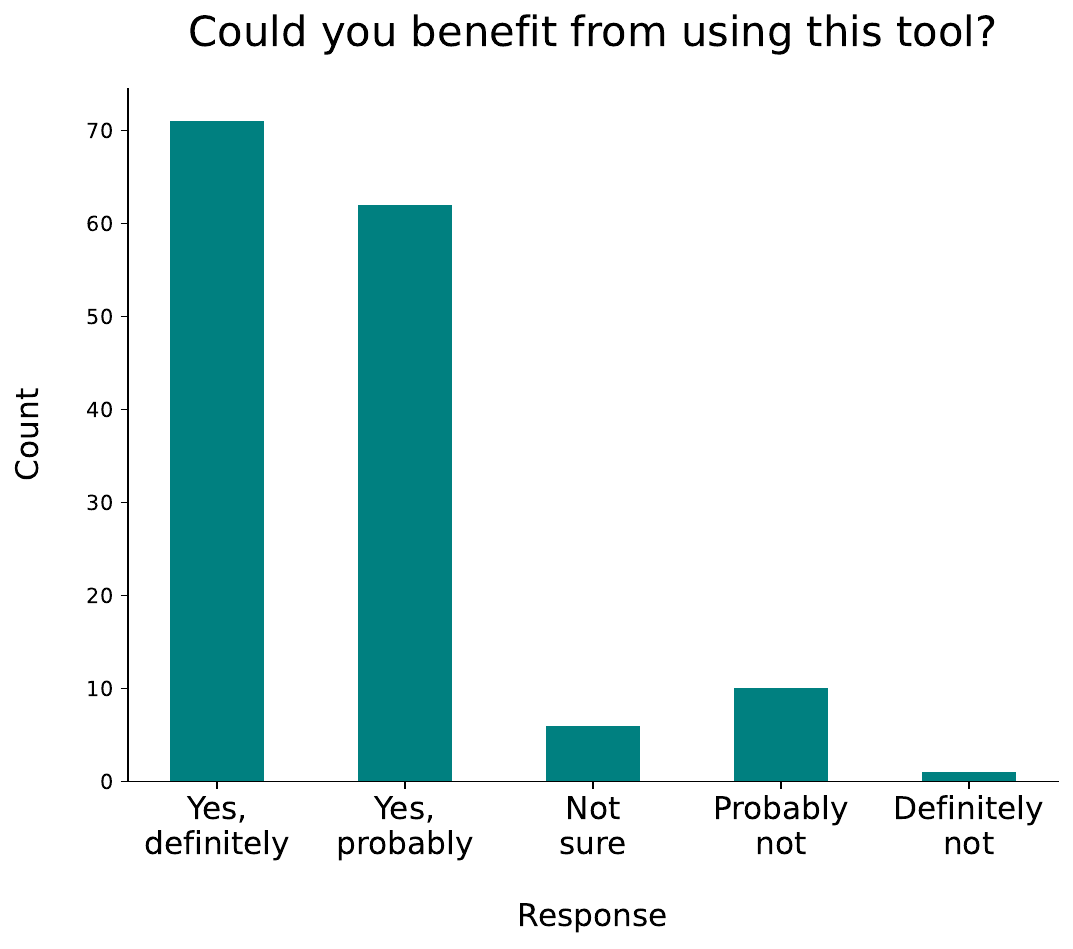}
        \caption{Personal benefit.}
        \label{fig:audience-subfig1}
    \end{subfigure}
    \hfill
    \begin{subfigure}{0.45\textwidth}
        \centering
        \includegraphics[width=\linewidth]{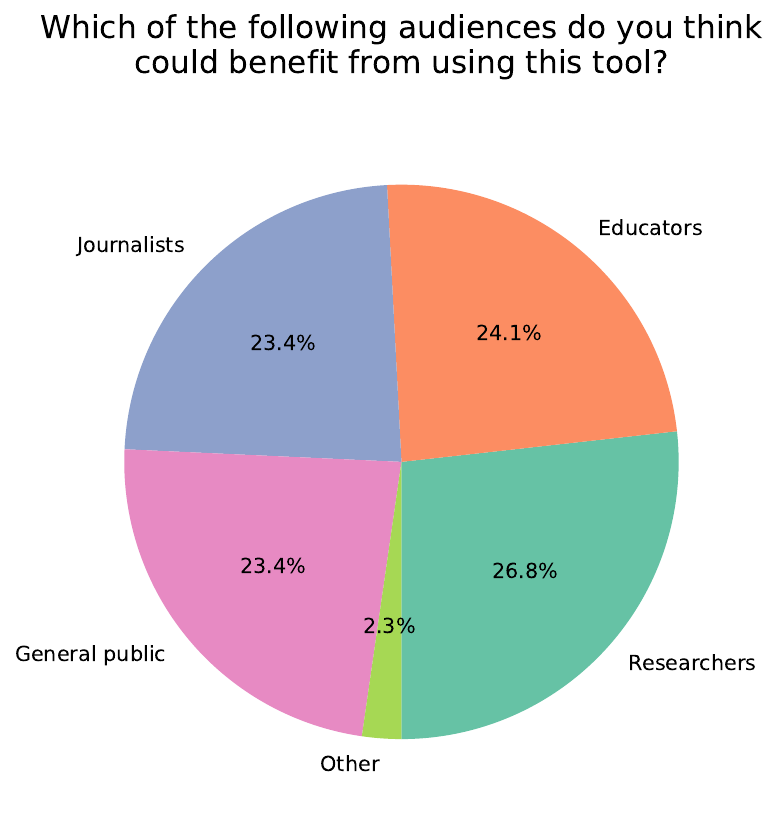}
        \caption{Recommended audience.}
        \label{fig:audience-subfig2}
    \end{subfigure}
    \caption{A vast majority of participants said that they would benefit from using this tool and considered both experts and the general public to be suitable audiences for it.}
    \label{fig:audience}
\end{figure}

In the user study (see Section \ref{sec:who_is_user}), we asked experts who they believed would benefit from the \mbd. Most suggested that, given their own extensive experience with media bias, the target audience might be everyday news consumers who lack similar training and expertise. Here, we repeated this question to ordinary people and asked who they thought would be the right audience for this tool (Figure \ref{fig:audience}). The responses were evenly distributed across all options, with participants viewing researchers, educators, journalists, and the general public as equally appropriate audiences for the tool. This highlights the broad appeal of the \mbd\ and shows that people in general consider this to be a useful tool for a wide range of people. 

\begin{figure}[htbp]
    \centering
    \begin{subfigure}{0.45\textwidth}
        \centering
        \includegraphics[width=\linewidth]{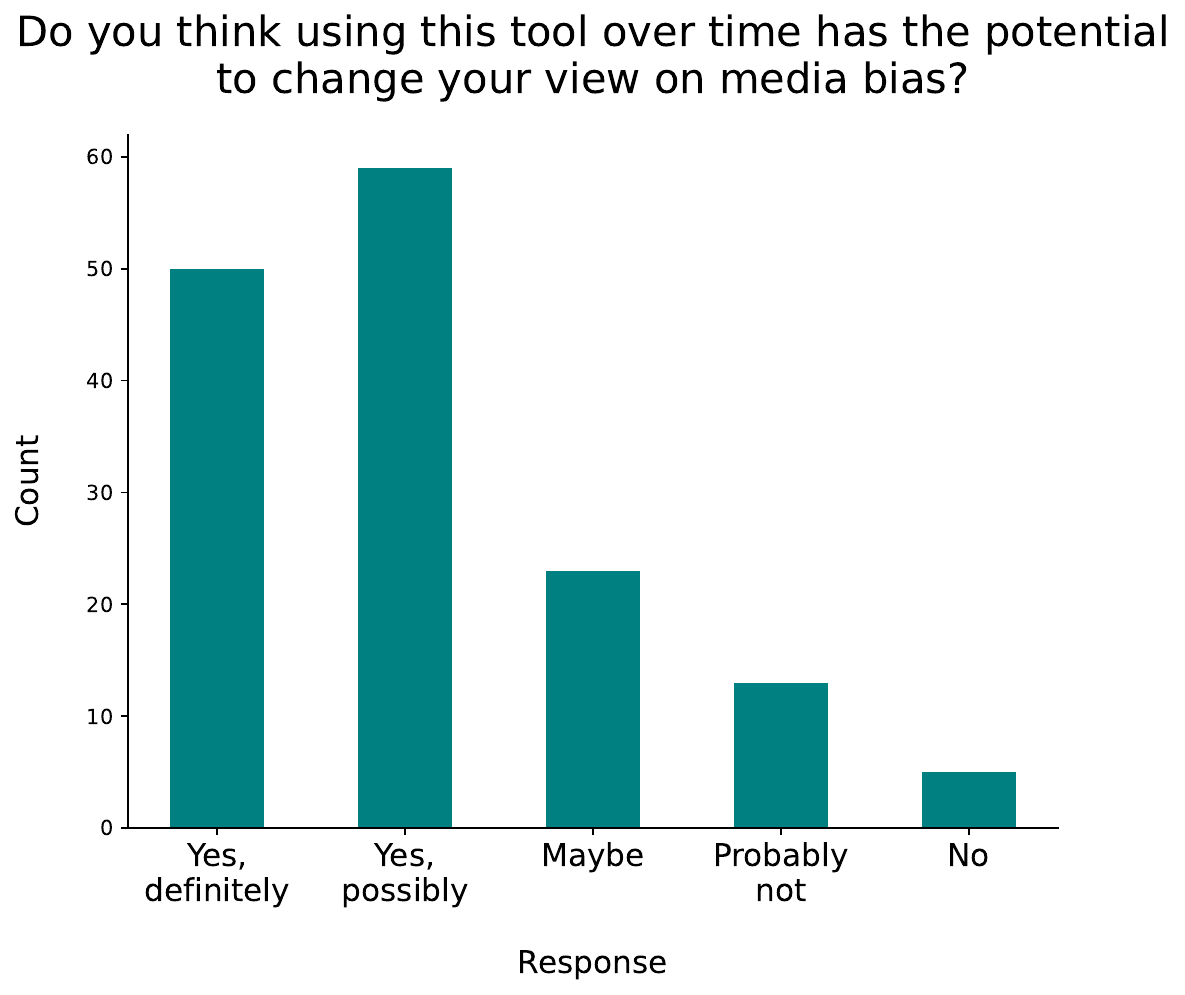}
    \caption{Impact of tool use over time.}
    \label{fig:overtimea}
    \end{subfigure}
    \hfill
    \begin{subfigure}{0.45\textwidth}
        \centering
        \includegraphics[width=\linewidth]{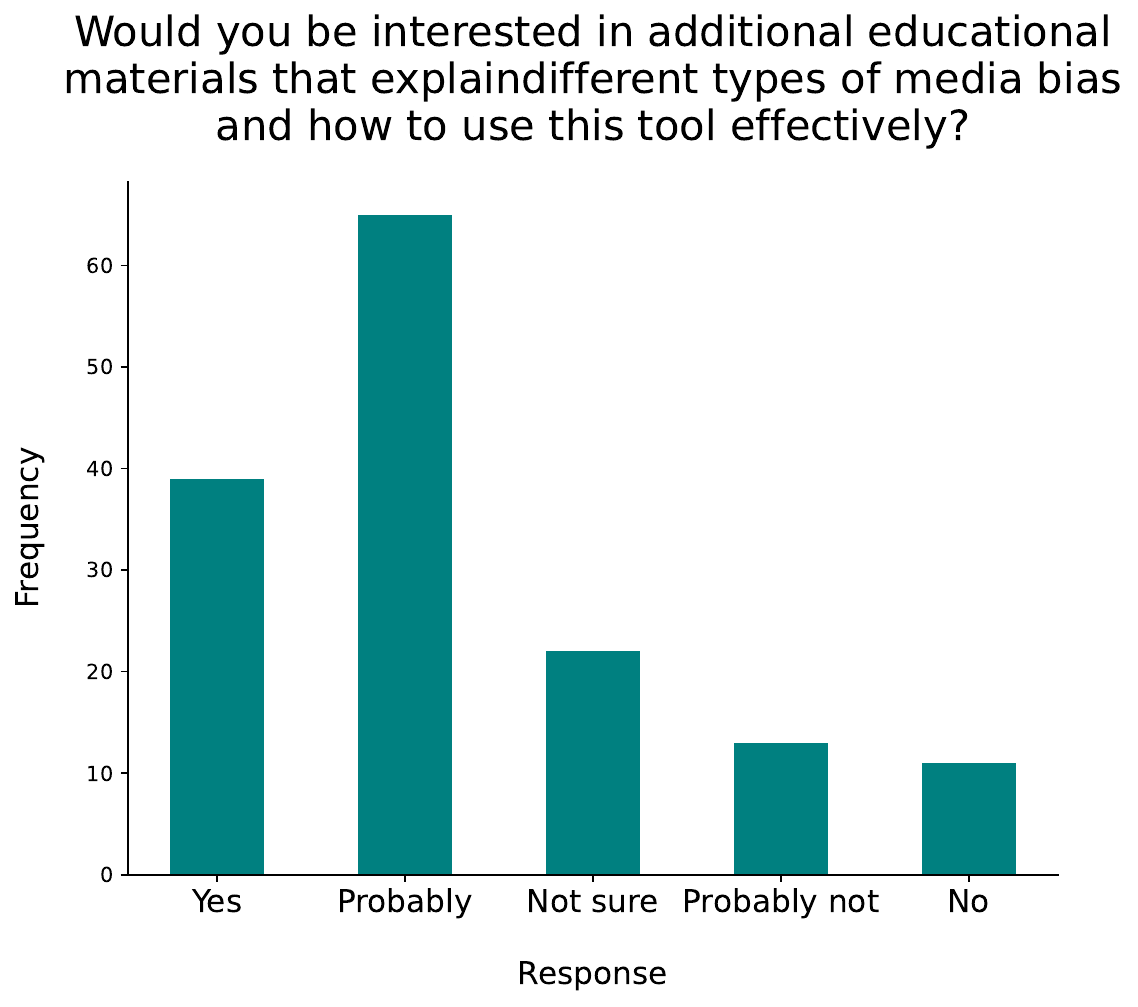}
        \caption{Usefulness of education materials to aid in tool usage.}
        \label{fig:overtimeb}
    \end{subfigure}
    \caption{A majority of participants said the \mbd\ had the potential to change their views on media bias over time. They also stated that they would find educational materials explaining different kinds of media bias helpful for getting the most out of this tool.}
    \label{fig:overtime}
\end{figure}

One of the major design goals of the \mbd\ is to \textit{show}, not \textit{tell} people about media bias. This is because we believe helping people recognize and understand the various ways media can present a biased picture of the world is more valuable than simply assigning bias ratings to different publishers. Because of the nature of the problem this tool is addressing, we anticipated that the short-term impact of engagement with the tool would be limited. To evaluate this, we surveyed respondents about their views on selection and framing bias before and after using the tool. The results indicated that there was no significant change in users' perception of bias as a result of this brief engagement (see Appendix \ref{sec:app_study2}, Figure \ref{fig:null_selection_framing}).
However, the true impact of using the \mbd\ will be apparent over time, as people continue to use it in tandem with their usual media consumption. When asked whether they thought this tool had the potential to change their views on media bias over time, the majority said that this would `definitely' (50/150) or `possibly' (59/150) be the case (Figure \ref{fig:overtimea}). Furthermore, a fair number of participants (39/150) also believed that educational materials that explained different kinds of media bias and more information about using the tool would be helpful (Figure \ref{fig:overtimeb}), with one of them noting that \textit{"understanding media bias thoroughly and knowing how to navigate the tool would enhance its usability and reliability"}.

\section{Discussion}

In this paper, we have introduced the \mbd\ and described two mixed-methods studies in which we investigated users' perceptions of the tool as well as its impact on their ability to explore and quantify media bias in news coverage. Here, we summarize the key findings for each of the research questions articulated above and characterize the \mbd's contribution to the existing state of media bias tools.

\subsection*{RQ1: Conveying Media Bias Through Political Lean and Tone}

The \mbd\ aims to empower people to understand media bias themselves instead of simply telling them which publishers are biased and which aren't---a value judgment that is subjective in and of itself. To achieve this, we designed the \mbd\ to be easily accessible to researchers, journalists, and everyday news consumers because media bias is a universal problem that affects people from all spheres. In our evaluation with media experts, we found that participants valued the affordances provided by the tool to examine and evaluate data beyond political lean, an important design consideration highlighted in \textbf{D1} (see Section \ref{sec:design_considerations_d1}). Participants also engaged in-depth with the topics they evaluated across lean and tone, as well as the events they viewed across publishers. Users dug deeper into the data to better understand common patterns across and within publishers, as described in \textbf{D2} (see Section \ref{sec:design_considerations_d2}).

Other tools such as AllSides and Media Bias/Fact Check consider lean at the publisher level, assigning a single label to a publisher across the political spectrum. Although this approach simplifies the problem greatly, it fails to capture the heterogeneous patterns of bias that can vary within a single publisher across different topics, events, or article types. For instance, an article's political lean or tone may shift depending on whether it is a news report, news analysis, or opinion piece, reflecting the organization’s priorities and context. It may also vary due to the evolving nature of publishers’ perspectives on emerging issues, such as elections or the COVID-19 pandemic. In contrast, the \mbd\ assigns labels at the article level, allowing for a more nuanced understanding of these within-publisher variations. Then, the tool aggregates this data at topic, subtopic, and publisher levels to provide high-level assessments while accounting for heterogeneous effects between articles.

We also note that while some tools attempt to evaluate publisher level reliability, we have avoided making article level reliability judgments. Assessing reliability at the article level is particularly challenging because automating these metrics requires contextual understanding beyond the text, which remains difficult for both LLMs and humans alike. The dynamic nature of news reporting, including the ability to correct errors post-publication, further complicates reliability assessment at the article level and presents an intriguing avenue for future research.

From a user experience perspective, we chose to distill complex information into two salient outcome measures that vary significantly by article within publisher: lean and tone, where we emphasize that tone is unrepresented in existing bias tools but is increasingly considered important to audience perception  \cite{robertson2023negativity}. The \mbd\ focuses on these two dimensions to provide an accessible, user-friendly starting point before expanding to additional variables. Several participants noted that the tool's interactivity gave users more autonomy to form their own judgments. Participants appreciated the dual focus on lean and tone, as well as the affordances and customization options offered by the tool. However, some participants suggested that simplifying the tool could improve usability, which implied that widening the scope of the \mbd\ to additional dimensions could introduce too much complexity in learning how to use it. These findings demonstrate that while deconstructing the multidimensional aspects of media bias is both feasible and impactful, user-centered design is crucial. The \mbd\ contributes a design-centered approach to media bias tools, focusing on interactive evaluation across multiple perspectives, within categories, and while comparing sources to each other.

\subsection*{RQ2: AI Skepticism and Human-in-the-Loop Design}

Our findings reveal a divergence in trust toward AI-driven media bias tools, with experts largely expressing skepticism and everyday users showing greater initial confidence in LLM-generated outputs. This skepticism among experts highlights the need for transparency, particularly around the LLM’s methodology and human-in-the-loop processes. Participants suggested adding transparency features, such as LLM explanations of article labels and examples of labeled articles, to help foster trust in the AI model’s decision-making process.

Our approach includes humans in the loop to validate the model's outputs and ensure they align with human judgment. This hybrid system balances automation with accountability, maintaining consistency without placing sole reliance on automated processes. Encouragingly, both experts and everyday news consumers generally showed increased trust in the tool after learning about the human-in-the-loop process. This finding is consistent with prior research showing that trust in AI grows when people know it has higher accuracy \cite{zhang_accuracy}. In our case, the humans-in-the-loop act as accuracy validators by ensuring the model’s outputs are reasonable and reliable.

As LLMs become increasingly popular, their role in shaping the information ecosystem grows more significant. Media bias tools like AllSides and Ground News are also incorporating AI into their features. While these websites are integrating AI to scale media bias assessments traditionally done by humans, their lack of transparency on both human and AI bias assessments leaves room for skepticism. Unlike these commercial tools, the \mbd\ is free and open source, with transparent prompts, methodology, and weekly human validation. We implement these safeguards to minimize the risks associated with unchecked automation and ensure the LLM is performing consistently.

Beyond immediate trust, our findings highlight broader issues associated with human-AI tool development. While experts who work closely with language models and understand their faulty nature remain skeptical, regular users are quickly adjusting to the new normal with LLMs. Prior research has shown that trust in AI grows with familiarity \cite{yu_user_trust}. Assuming the research community continues to validate these models and ensure they remain calibrated, the trust growing among users is encouraging. However, it also underscores the critical need to design tools that keep humans in the loop, prioritize transparency, and incorporate safeguards to address initial skepticism—preventing potential failures before it’s too late.

\subsection*{RQ3: Contributing a Practical Tool to the Media and Communications Research Community}

Our evaluations across different user groups showed that the \mbd\ effectively serves a diverse audience. While media experts predominantly suggested that the tool was more relevant to everyday news consumers, those news consumers felt that not only they, but also experts---journalists, educators, and researchers---could benefit from using this tool in the long run. Participants across user groups highlighted the \mbd’s value in revealing unexpected patterns, such as selectivity in covering major events and overall political lean of topics like the economy. The guided exploration tasks in the follow-up survey found that the tool not only confirmed prior beliefs but also introduced new insights, showing that users critically engaged with the tool.

We also discovered that news consumers saw the \mbd\ as a tool that could shift their views on media bias over time, with many expressing interest in additional educational resources. This suggests that the tool not only revealed biases they hadn't noticed before but also sparked curiosity to learn more. In our follow-up study, a significant number of participants indicated they would benefit from extra guidance on using the tool effectively. This suggests that incorporating more resources on communication techniques like prebunking \cite{amazeen2022cutting}, accuracy nudges and fact-checking \cite{pennycook2022nudging}, and debunking \cite{chan2017debunking} could supplement the \mbd\ and further support the learning process. Although accuracy nudges and fact-checking are more pertinent to social media content and may not be directly applicable to mainstream media outlets—which typically do not contain outright factual inaccuracies—parallel strategies could be developed to prompt news consumers to recognize subtle biases, such as those involving framing and selection. The \mbd\ is purposefully designed to be modular, prioritizing popular news publishers with the greatest reach \cite{allen2024quantifying}. By focusing on high-impact publishers, we can sustainably validate LLMs’ outputs and implement necessary safeguards before expanding the tool further.

The true impact of the \mbd\ will become apparent as people regularly view the data it presents and incorporate it into their daily lives. But beyond individual learning, the \mbd\ contributes to democratic engagement by making media bias more transparent and providing people with resources to contextualize narratives and question editorial choices. Our hope is that equipping users with the tools and knowledge to recognize instances of bias will help the public benefit in the long run. These studies have helped us more clearly recognize that the \mbd's primary impact lies in improving users' understanding of media bias over time rather than simply showing them who is biased and who is not. In the wake of the 2024 election, where misinformation, one-sided narratives, and concerns about election integrity dominated public discourse, our findings show that interactive bias detection tools may play an important role in contextualizing what happens in the world during the digital age. 

\section{Limitations and Future Work}
\label{sec:limitations}

Our study helped identify areas for improvement and potential features to incorporate into future work. Below, we detail these limitations, structured across design, data, and methodological considerations.

\subsection{Design-Related Limitations}

Media experts and survey participants provided valuable feedback on the need for usability enhancements. Several participants requested a search bar to help users find specific topics and subtopics. Similarly, the Events dashboard requires users to know the exact date of an event to access related articles, which can be challenging when they only remember general time frames or keywords. Implementing a search feature for different events would improve usability by allowing users to find relevant content more reliably. Another recommended feature was the ability to click on URLs that go directly to the publisher’s website to view the original articles. While this feature is currently implemented on the Events dashboard, where users can read and verify the top facts themselves, participants suggested that extending the functionality to the Coverage dashboard could increase user engagement and build trust in the platform. Another feature to build trust we plan to provide involves showing examples of articles in each tone bucket (e.g. Democrat to Republican) and lean bucket (e.g. Very Negative to Very Positive) as users focus on a specific topic or subtopic. Addressing these issues will help users access familiar news stories, making it easier to start their exploration on the \mbd.

Additionally, while the Events dashboard presents a qualitative comparison of coverage differences between publishers---such as fact selection and framing---users requested aggregated insights, such as topic, tone, and lean labels to help them quickly identify these differences. In response, we plan to assign topic and subtopic labels to each event based on the majority class of their constituent articles and display these labels with aggregated tone and lean data directly on the Events page. These updates will enable quicker comparisons and a more comprehensive exploration of coverage differences.

Finally, many participants highlighted the tool’s potential as an educational resource for fostering media literacy. To build on this, we aim to add a section to our website that can guide users who are less familiar with media bias. This section may include blog posts (where team members write analyses using the \mbd), videos, and interactive guides explaining how to use the \mbd\ to identify different forms of media bias (e.g., selection, framing, tone). These resources aim to empower users to better navigate not only the \mbd\, but the information ecosystem as a whole.

\subsection{Data-Related Limitations}

The \mbd\ currently tracks only ten publishers and collects their top 20 articles in every interval. While this covers the most prominent news published by the most widely-read publishers, it excludes coverage from local or more niche sources. Several participants highlighted that this approach misses other perspectives in the news. We built the dashboard with ten publishers, due to the significant financial and engineering costs associated with data collection and labeling pipelines. Each publisher requires a tailored data collection pipeline which involves manual text preprocessing. Additionally, using state-of-the-art, closed-source models (i.e., OpenAI’s GPT-4o) on a daily basis incurs significant costs that scale linearly with the number of publishers analyzed. To select these ten publishers, we sought out a set of sources that are popular among news readers \cite{statista_news_sites}, while also maintaining diversity across the political spectrum (we added Breitbart for its agenda-setting power on the right, The Guardian and Huffpost on the left, and the Wall Street Journal for its financial leadership). We are currently in the process of expanding our data collection pipeline to cover the top 30 stories per news publisher and plan to expand our list of publishers beyond legacy media. 

Another limitation of our tool is that the topic and subtopic lists are driven by what is actively discussed in the news. When the human-in-the-loop annotators identify major themes that have emerged in the news, we make updates to the topic list. This creates a blind spot for important issues that are not being reported by research assistants, as they may never appear in our analysis. Additionally, our current approach assigns each article to a single topic and subtopic. This rigid assignment of topics and subtopics can overlook articles that fall under multiple options. Not including them in one of those options affects what we see in the coverage numbers and may not fully reflect reality. This highlights the broader challenge of classifying a complex news article into a finite number of categories. In future iterations of the \mbd, we may explore more flexible methods to accommodate overlapping themes. For example, allowing a single article to be associated with "most likely" topics or subtopics or adjusting prompts to support multiclass classification.

\subsection{Limitations of LLMs in Media and Politics}

Our tool relies on instruction-tuned, pre-trained LLMs to generalize tasks in a zero-shot setting, which introduces inherent challenges \cite{zhang2023instruction}. While LLMs are efficient for text classification, they are not immune to biases present in their training data, which can subtly influence downstream tasks such as political lean labeling \cite{feng-etal-2023-pretraining}. This has important implications for the \mbd’s results, particularly when analyzing political news topics. When prompting the LLM for explanations on its political lean label, we can identify that they draw conclusions from not only the article text itself, but from context and inferences from its training data. This results in the LLM inherently associating certain topics with Democratic leanings (such as climate change or women’s healthcare), and others with Republican leanings (such as immigration or crime), regardless of the specific arguments presented in the article. We must be cautious as the nature of these contexts and inferences may shift over time.

When measured using both novel and established methods to assess political bias, GPT model variants, including GPT-4 and GPT-3.5-Turbo, tend to align with a left-leaning political ideology \cite{feng-etal-2023-pretraining, rutinowski2023self, motoki2024more}. This may relate to the model’s tendency to align certain issues with the Democratic or Republican party’s political agenda. While individual political lean labels could be subject to LLM biases, their relative values, and the differences between labels on articles about the same event, are informative. Our research is grounded in comparative analysis between publishers on multiple levels, and we intentionally provide granularity in all our analyses to allow users to draw their own conclusions. Furthermore, the \mbd\ emphasizes features like tone, news topics, and event coverage, which are less susceptible to political bias in LLMs compared to political lean labels.

\section{Conclusion}

We believe that helping people identify and quantify bias across different stories and publishers, along with exposing readers to alternative narratives, can enhance public awareness of the editorial choices made by news publishers. From this perspective, the \mbd\ could help counter the potential for creating separate realities by empowering users to independently compare how different publishers cover the same topics or events \cite{bram2024beyond}. For instance, participants in our user study noted that the Events dashboard helped them see discrepancies in coverage of major events, such as the Gaza ceasefire negotiations, where some outlets prioritized domestic political stories instead. Additionally, everyday users reported that the ability to drill down into subtopics and compare political lean and tone encouraged them to challenge their preconceptions about specific publishers. By providing a high-level view of the media landscape, the \mbd\ can help users engage with diverse perspectives and hold media outlets accountable for their power to shape public perception. Moreover, the tool may also serve as a valuable resource for researchers, educators, and journalists, enabling them to check their own biases and quantitatively assess media coverage.

We close by re-emphasizing that the goal of the \mbd\ is to empower users to discover bias on their own, not to tell them what is biased and what is not. Our focus is on understanding how news publishers' decisions influence public perception, and how emerging narratives are shaped by the emphasis placed on them by different mainstream media outlets. This exploration is not about finding fault, but about analyzing how the natural processes of news selection and framing influence the stories we encounter in our everyday lives \cite{Watts_etal}. We therefore hope our work contributes to meaningful discussions at the intersection of media, the public, and politics, by promoting critical thinking about how news is produced, consumed, and absorbed, and ultimately the worldviews and choices it shapes.

\begin{acks}
We thank Polygraph for their support in developing the front end, Ajay Patel and Bryan Li for helpful discussions on the methodology, and Harsh Kumar for encouraging us to conduct user studies on the Media Bias Detector as well as present it to a HCI audience. DJW and DR acknowledge their long collaboration with Markus Mobius, during which they worked on pre-LLM methods to cluster news articles and display them visually on a web-based dashboard as part of what was called “Project Ratio,” initially at Harmony Labs and subsequently at Microsoft Research (\href{https://www.microsoft.com/en-us/research/project/project-ratio/}{https://www.microsoft.com/en-us/research/project/project-ratio/}). This research was developed with funding from Richard Jay Mack and the Defense Advanced Research Projects Agency’s (DARPA) SciFy program (Agreement No. HR00112520300). The views expressed are those of the authors and do not reflect the official policy or position of the Department of Defense or the U.S. Government.
\end{acks}

\bibliographystyle{ACM-Reference-Format}
\bibliography{references}

\clearpage


\appendix

\section{Appendix}
\subsection{User Study with Media Experts}
\label{sec:app_study1}

\renewcommand{\thefigure}{A\arabic{figure}}
\setcounter{figure}{0}

\subsubsection*{Task Scenario}
\label{sec:apdx_task}

The same two scenarios were presented sequentially to the participants:

\begin{figure}[h]
    \centering
    \includegraphics[width=\linewidth]{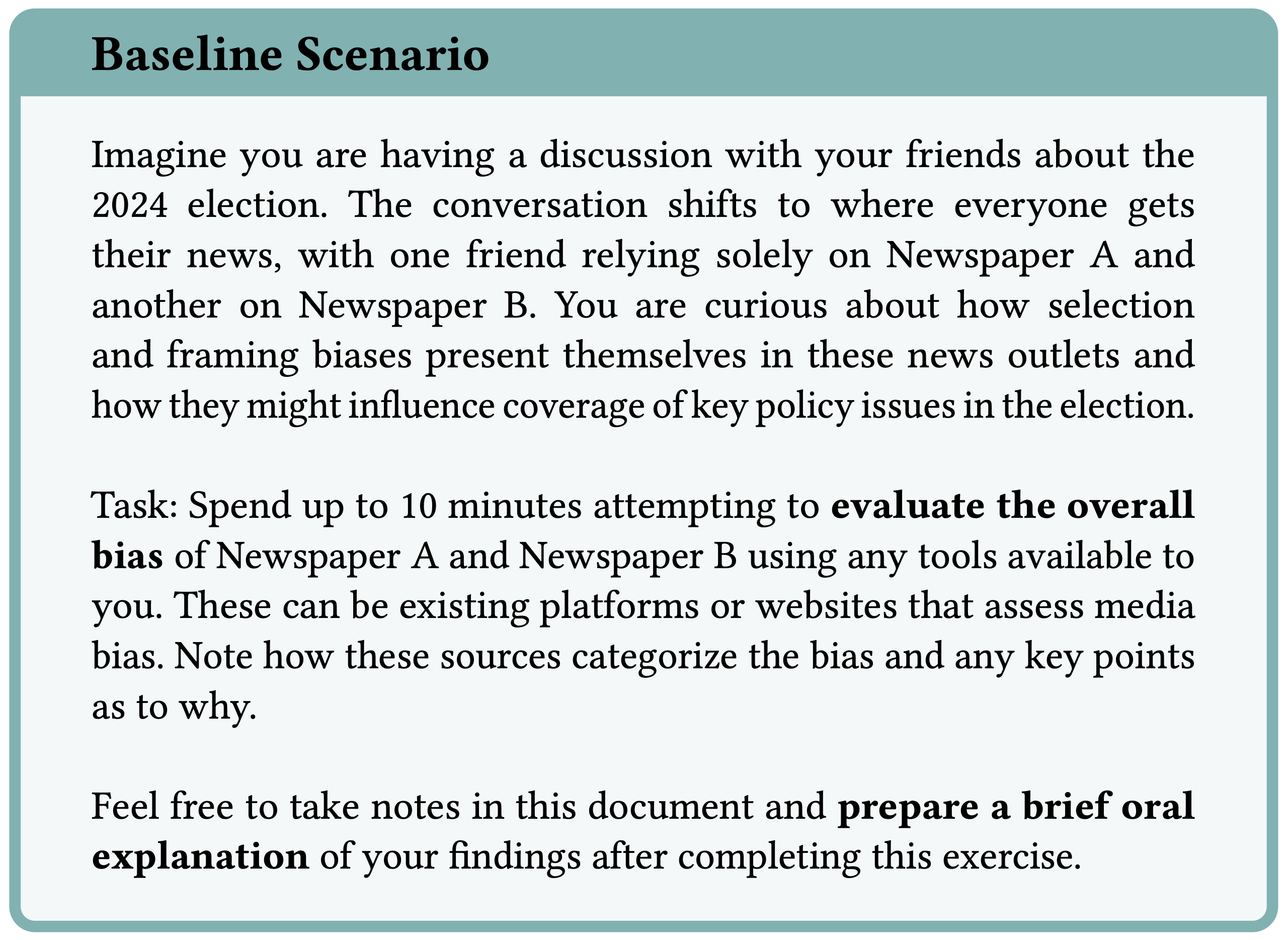}
\end{figure}

\begin{figure}[h]
    \centering
    \includegraphics[width=\linewidth]{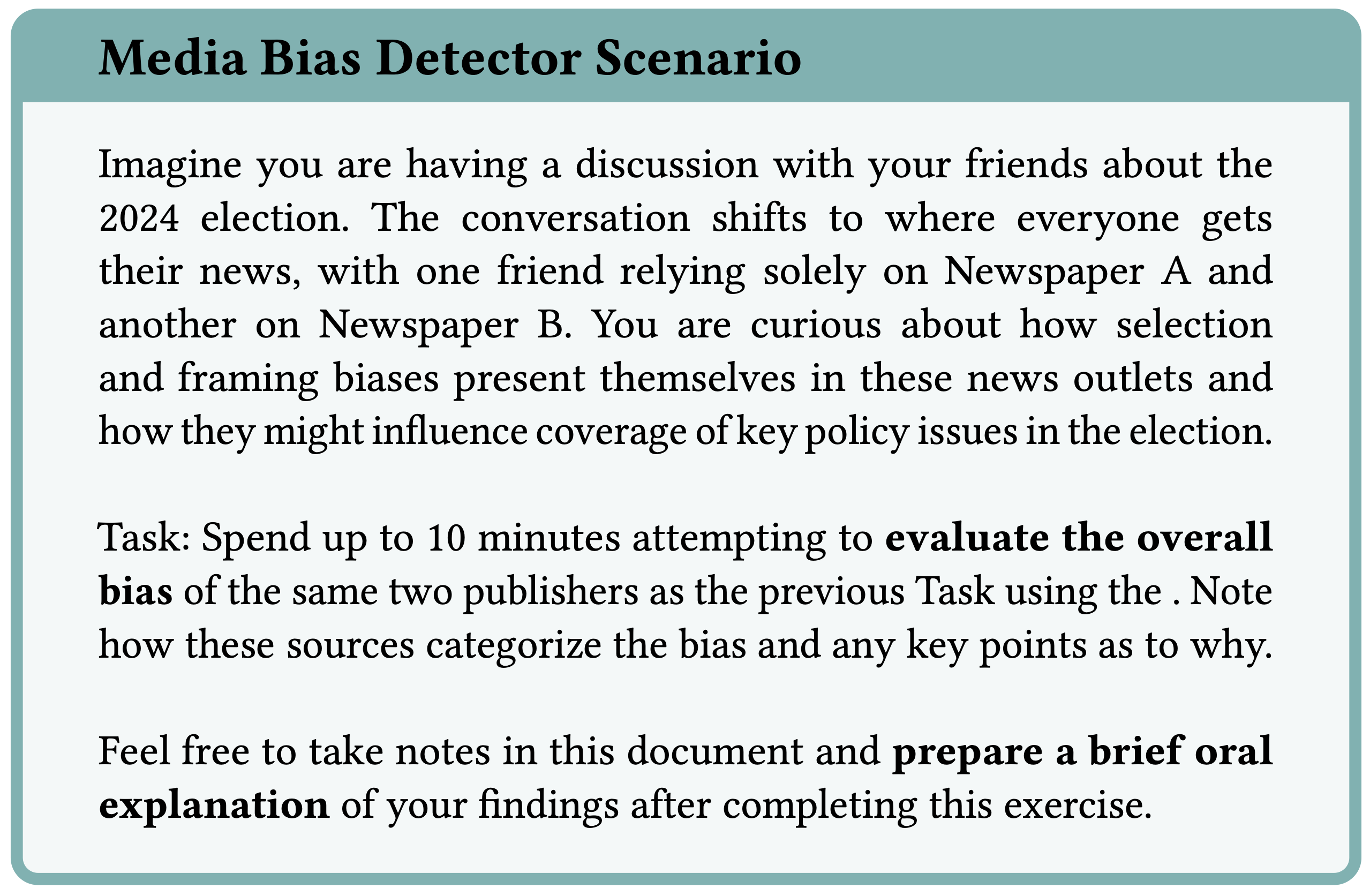}
\end{figure}

\subsubsection*{Guiding Questions from Semi-Structured Discussion}
\label{appendix:guiding_questions}

These guiding questions were developed to steer the discussion, though not all were necessarily used during the interviews:
\newline

\textbf{Pre-Task}
    \begin{itemize}
        \item Before we get started, can you tell me more about what got you interested in [media and journalism] OR [political science] OR [AI \& society]?
        \item What does journalism mean to you? What is its purpose?
        \item In your own words, what does media bias mean to you? 
        \item How do you recognize bias when reading news? What specific signs do you look for?
        \item How do you see the role of technology, such as AI and large language models, affecting the study and mitigation of media bias?
    \end{itemize}

\textbf{Post-Task}
    \begin{itemize}
        \item How could this tool be used by you or others in your industry?
        \item What were features that you found useful and what were features you disliked?
        \item If you could make a magical tool that does anything you want, what features would it have? Is there anything that you really wish existing systems could do? 
    \end{itemize}

\subsubsection*{Survey Questions}
\label{appendix:survey_questions}

We asked participants for demographic information and evaluations for the baseline and \mbd\ tools through Qualtrics surveys. The surveys are viewable through Open Science Framework (OSF). 

\begin{itemize}
    \item \href{https://osf.io/cmtkd?view_only=6dcbdd78e5d940c8bbabc2378e4bd1ed}{Pre-Survey Demographic Questions}
    \item \href{https://osf.io/82cbt/?view_only=6dcbdd78e5d940c8bbabc2378e4bd1ed}{Baseline Evaluation}
    \item \href{https://osf.io/x4vp9?view_only=6dcbdd78e5d940c8bbabc2378e4bd1ed}{Tool Evaluation}

\end{itemize}

\subsection{Follow-Up Survey of News Consumers}
\label{sec:app_study2}

\begin{figure*}
    \centering
    \includegraphics[width=0.6\linewidth]{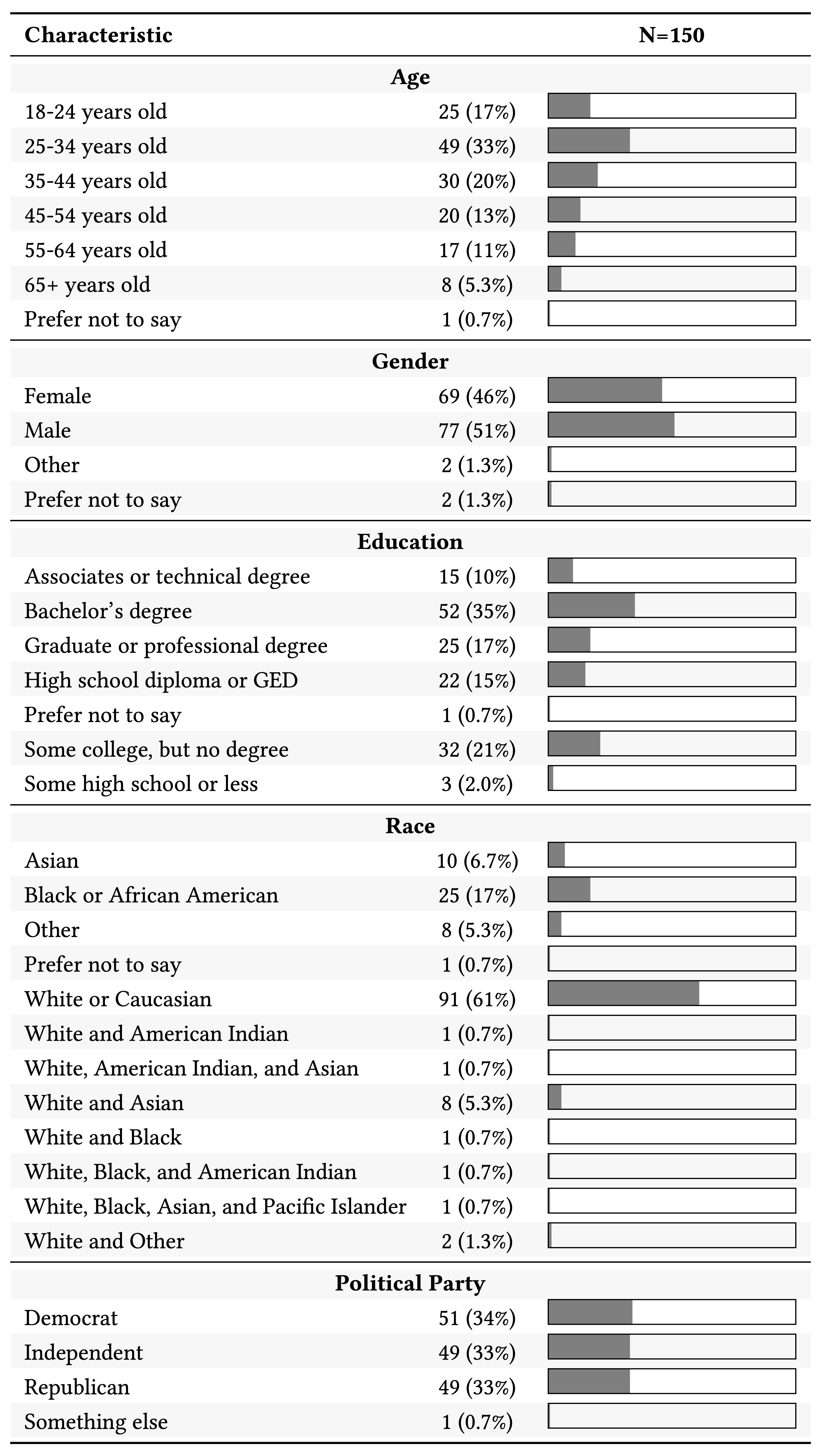}
    \caption{Characteristics of the Follow-Up Survey Population.}
    \label{apx:population}
\end{figure*}

The follow-up survey on everyday news consumers was conducted via a Qualtrics survey. Figure \ref{apx:population} provides a demographic breakdown of the 150 participants. The survey is viewable through OSF: 
\begin{itemize}
    \item \href{https://osf.io/yzw7d?view_only=6dcbdd78e5d940c8bbabc2378e4bd1ed}{News Consumers Survey}
\end{itemize}

\begin{figure}[h!]
    \centering
    \includegraphics[width=\linewidth]{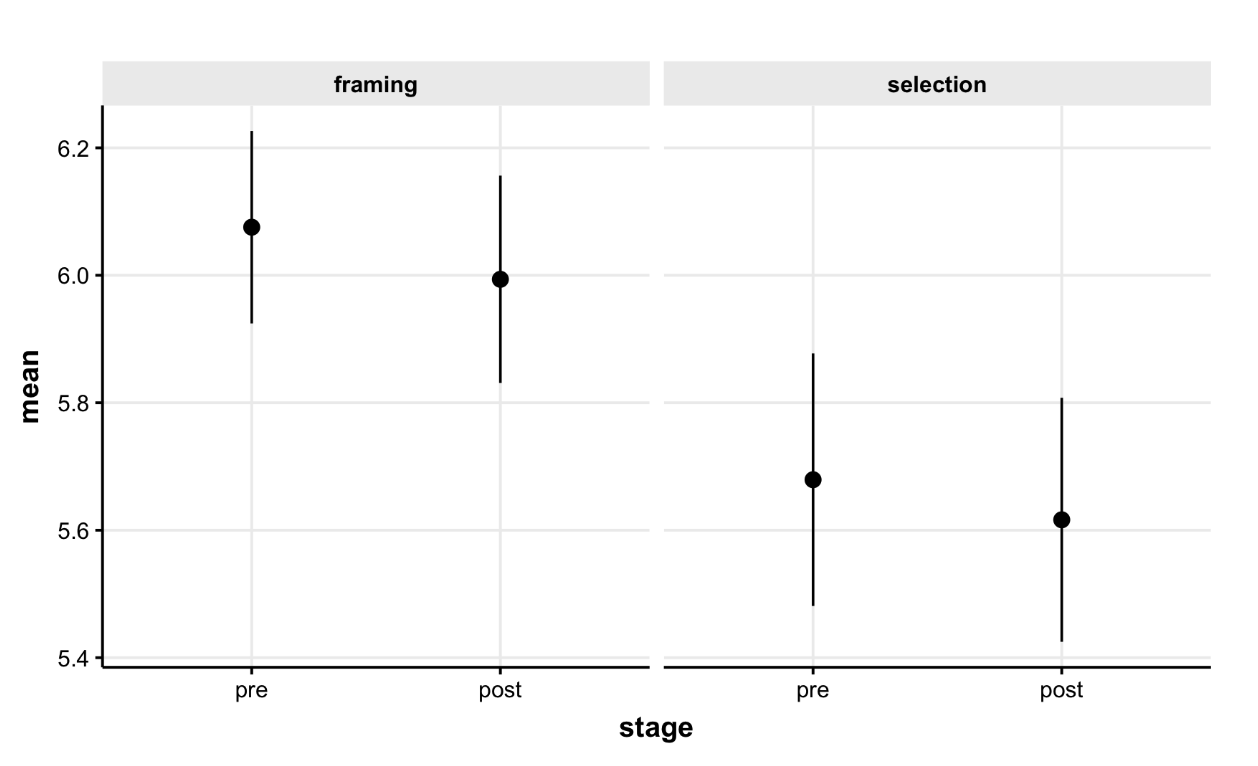}
    \caption{A comparison of pre- and post-task ratings by participants on their perception of selection and framing bias in the news.}
    \label{fig:null_selection_framing}
\end{figure}

\subsection{Master Topic List}
\label{sec:app_topics}

\begin{figure*}[h]
    \centering
    \rotatebox{90}{\includegraphics[trim=0 3cm 0 0, clip, width=1.1\textwidth]{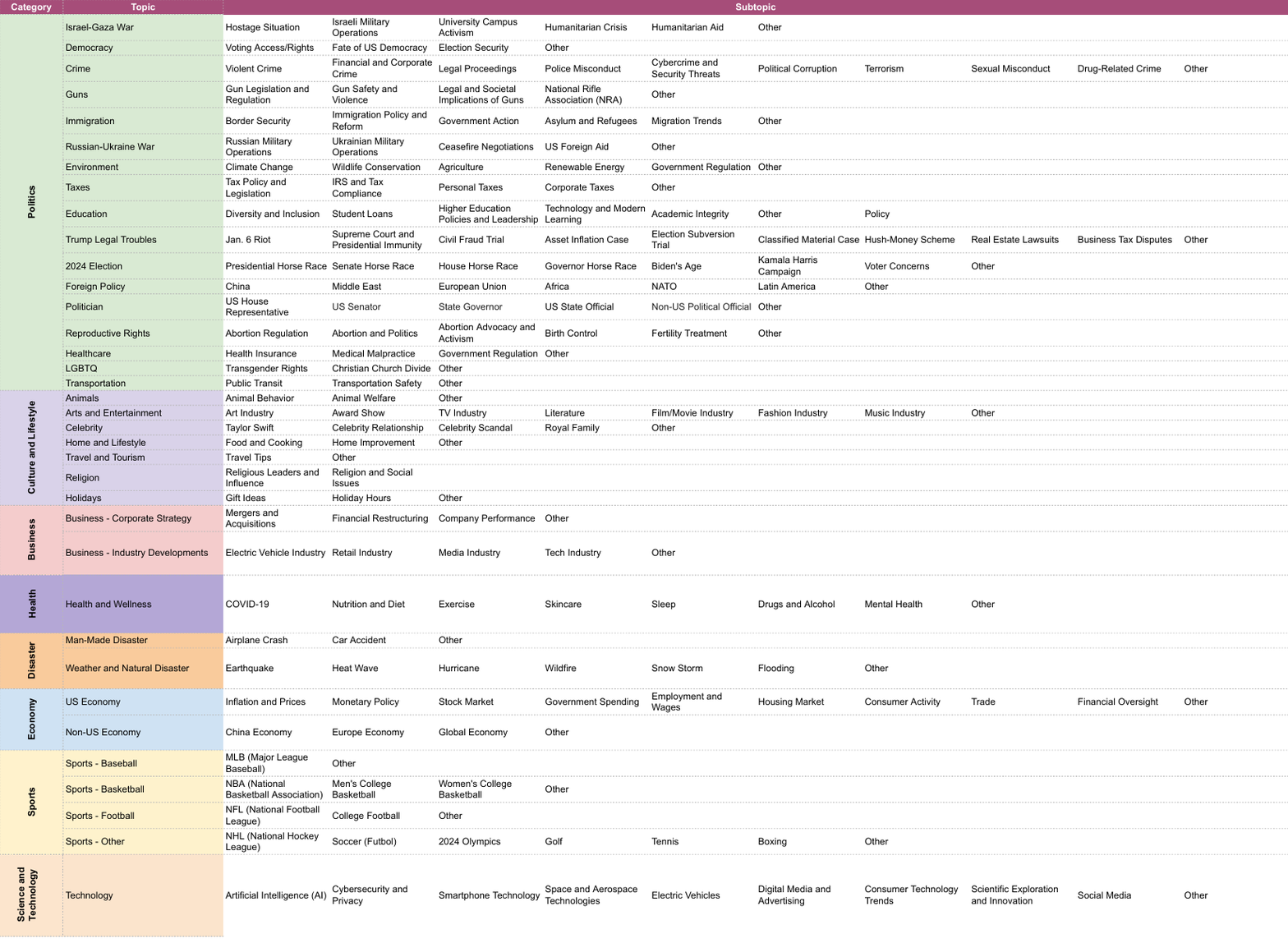}}
    \caption{Current News Category, Topic, and Subtopic Hierarchy.}
    \label{fig:topics}
\end{figure*}

See Figure \ref{fig:topics}.

\subsection{Detailed Methodology}
\label{sec:website_methodology}

Visit the Media Bias Detector website's \href{https://mediabiasdetector.seas.upenn.edu/methodology/}{\emph{Methodology section}} for the most up-to-date details. The methodology is also available in PDF form in the supplementary materials.

\end{document}